\documentclass[aps,prl,floatfix,twocolumn,showpacs]{revtex4}%
\usepackage{amsmath}
\usepackage{graphicx}
\usepackage{amsfonts}
\usepackage{amssymb}%
\providecommand{\U}[1]{\protect\rule{.1in}{.1in}}
\providecommand{\U}[1]{\protect\rule{.1in}{.1in}}
\providecommand{\U}[1]{\protect\rule{.1in}{.1in}}
\providecommand{\U}[1]{\protect\rule{.1in}{.1in}}
\providecommand{\U}[1]{\protect\rule{.1in}{.1in}}
\providecommand{\U}[1]{\protect\rule{.1in}{.1in}}
\providecommand{\U}[1]{\protect\rule{.1in}{.1in}}
\providecommand{\U}[1]{\protect\rule{.1in}{.1in}}
\providecommand{\U}[1]{\protect\rule{.1in}{.1in}}
\providecommand{\U}[1]{\protect\rule{.1in}{.1in}}
\begin{document}
\title{Microwave spectroscopy of a Cooper pair beam splitter}
\date{\today}

\author{Audrey Cottet}

\affiliation{Laboratoire Pierre Aigrain, Ecole Normale Sup\'{e}rieure, CNRS UMR 8551,
Laboratoire associ\'{e} aux universit\'{e}s Pierre et Marie Curie et Denis
Diderot, 24, rue Lhomond, 75231 Paris Cedex 05, France}

\begin{abstract}

This article discusses how to demonstrate the entanglement of the split Cooper pairs
produced in a double-quantum-dot based Cooper pair beam splitter (CPS), by performing
the microwave spectroscopy of the CPS. More precisely, one can study the DC current response of such a CPS to two on-phase microwave gate irradiations applied to the two CPS dots. Some of the current peaks caused by the microwaves show a strongly nonmonotonic variation with the amplitude of the irradiation applied individually to one dot. This
effect is directly due to a subradiance property caused by the coherence of the split pairs. Using realistic parameters, one finds that this effect has a measurable amplitude.

\end{abstract}
\maketitle

\section{I. Introduction}

Quantum entanglement between spatially separated particles represents a
promising resource in the field of quantum computation and communication.
However, this fascinating behavior can be difficult to observe in practice due
to decoherence caused by the particles environment. This is why the "spooky
action at a distance" was first demonstrated with photons, atoms, or ions
which can be naturally placed in weakly interacting
conditions\cite{Aspect,atoms,ions}.

Observing electronic entanglement in solid state systems is a-priori more
challenging since an electronic fluid is characterized by a complex many-body
state in general. However, quantum entanglement has been recently observed on
superconducting chips\cite{Supraqubits}. In this case, the particles are
replaced by superconducting quantum bits, which can be sufficiently well
isolated from the outside world thanks to the rigidity of the superconducting
phase, if an appropriate circuit design is used. In these experiments, the
entangled degrees of freedom are defined from the charges of small
superconducting islands, or from the persistent current states of a
superconducting loop, for instance\cite{You}.

Superconductors enclose another natural source of entanglement which has not
been exploited so far, i.e. the spin entanglement of its Cooper pairs. In a
conventional superconductor, Cooper pairs gather two electrons correlated in a
spin-singlet state. The use of this resource for entanglement production
requires to build hybrid circuits in which the superconductors are connected
to non-superconducting elements which allow the spatial separation of Cooper
pairs. In principle, a double quantum dot circuit connected to a central
superconducting contact (input) and two outer normal metal contacts (outputs)
facilitates this process\cite{Recher:01}. Such a ''Cooper pair splitter''
(CPS) has been realized recently by using double dots formed inside
semiconducting nanowires \cite{Hofstetter:09,Hofstetter:11,Schindele} or
carbon nanotubes\cite{Herrmann:10, Herrmann:12}. The spatial splitting of the
Cooper pairs has been demonstrated from an analysis of the current response of
the CPS to a DC voltage bias. However, the spin entanglement of the split
pairs was not tested by these experiments.

It has been suggested to use the noise cross-correlations of the electrical
current to characterize the degree of entanglement of pairs of
electrons\cite{Martin:96,Anantram,Burkard:99,Lesovik,Borlin:02,Samuelsson:02,Chtchelkatchev,Sauret,Chevallier,Rech}%
. Alternatively, Ref. \cite{CKLY} proposes to put in evidence spin
entanglement by coupling the CPS to a microwave cavity. In this reference, a
double quantum dot formed inside a single wall carbon nanotube is considered.
Spin-orbit interaction produces a coupling between electronic spins and cavity
photons. Such a coupling leads to a lasing effect which involves a transition
between the spin singlet state in which Cooper pairs are injected and some
spin triplet states. This effect vanishes when the spin/photon coupling is
equal in the two dots, due to a subradiance property caused by the entangled
structure of the spin-singlets. However, realizing such an experimental scheme
is challenging since it requires to couple a complex quantum dot circuit to a
photonic cavity\cite{Delbecq,Frey,Xiang}.

The present work suggests an alternative strategy to exploit the subradiance
of spin-orbit induced transitions between spin singlet and spin triplet CPS
states. One can measure the DC current at the input of the CPS when microwave
gate voltage excitations are applied separately to the two CPS dots. The
microwave-induced state transitions mediated by spin-orbit coupling result in
current peaks at the input of the CPS versus the dots DC gate voltages.
Assuming that two on-phase microwave excitations are applied to the two dots,
these peaks vanish when the amplitude of the two excitations become equal.
This subradiant behavior is directly related to the spin-entanglement of the
split Cooper pairs hosted by the CPS.

This article is organized as follows. Section II defines the CPS hamiltonian,
for a single wall carbon nanotube based implementation. Section III discusses
the CPS even-charged eigenstates in the absence of the microwave excitations
and without the normal metal contacts. Section IV describes the coupling
between the CPS even-charged eigenstates and the microwave excitations.
Section V describes the CPS state dynamics in the presence of the
voltage-biased normal metal contacts, by using a master equation description.
Section VI describes the results given by this approach, an in particular the
predictions obtained for the DC current at the input of the CPS. Section VII
presents further examination and modifications of the model, which are useful
to put the results of section VI into perspective. In particular, it discusses
the role of atomic-scale disorder in the nanotube, the role of the form
assumed for the spin-orbit interaction term, and possible microwave induced
transitions in the CPS singly occupied charge sector. Section VIII compares
the measurement strategy discussed in this work to the one of Ref.
\cite{CKLY}. Section IX concludes. Although this article focuses on a
carbon-nanotube-based CPS, the entanglement detection scheme discussed in this
work could be generalized to other types of quantum dots with spin-orbit
coupling like e.g. InAs quantum dots, in principle.

\section{II. Hamiltonian of the CPS}

Let us consider the circuit represented schematically in Fig. 1. Two normal
metal contacts and a superconducting contact are used to define two quantum
dots $L$ and $R$ along a single wall carbon nanotube. The superconducting
contact is connected to ground, and a bias voltage $V_{b}$ is applied to the
two normal metal contacts. The dot $L(R)$ is connected capacitively to a DC
gate voltage source $V_{g}^{L(R)}$ and a microwave gate voltage source
$V_{ac}^{L(R)}(t)$. In the following, it is assumed that $V_{ac}^{L}(t)$ and
$V_{ac}^{R}(t)$ are in-phase, i.e. $V_{ac}^{L(R)}(t)=v_{ac}^{L(R)}\sin
(\omega_{RF}t)$. \begin{figure}[h]
\includegraphics[width=0.5\linewidth]{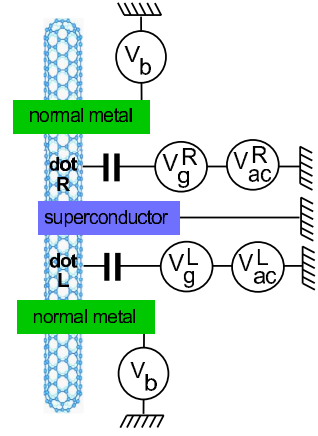}\caption{{}Scheme of a
Cooper pair splitter made out of a carbon nanotube. The two quantum dots $L$
and $R$ are defined by the normal metal contacts (in green) and the
superconducting contact (in blue) deposited on top of the carbon nanotube (in
light blue). The dot $L(R)$ is capacitively coupled to a DC gate voltage
$V_{g}^{L(R)}$ and microwave gate voltage $V_{ac}^{L(R)}$. The superconducting
contact is connected to ground and the normal metal contacts are biased with a
voltage $V_{b}$.}%
\end{figure}Inside the left and right dots $i\in\{L,R\}$, an electron with
spin $\sigma\in\{\uparrow,\downarrow\}$ can be in the orbital $\tau
\in\{K,K^{\prime}\}$ of the nanotube, which is reminiscent from the
$K/K^{\prime}$ degeneracy of graphene. One can use a double dot hamiltonian
which takes into account the proximity effect caused by the superconducting
contact, i.e.%

\begin{align}
H_{DQD}^{eff} &  =%
{\displaystyle\sum\limits_{i,\tau,\sigma}}
(\varepsilon+\Delta_{so}\tau\sigma)n_{i\tau\sigma}+H_{prox}\label{H}\\
&  +\Delta_{K\leftrightarrow K^{\prime}}%
{\displaystyle\sum\limits_{i,\sigma}}
(d_{iK\sigma}^{\dag}d_{iK^{\prime}\sigma}+d_{iK^{\prime}\sigma}^{\dag
}d_{iK\sigma})\nonumber\\
&  +t_{ee}%
{\displaystyle\sum\limits_{\tau,\sigma}}
(d_{L\tau\sigma}^{\dag}d_{R\tau\sigma}+d_{R\tau\sigma}^{\dag}d_{L\tau\sigma
})\nonumber
\end{align}
with
\begin{equation}
H_{prox}=t_{eh}%
{\displaystyle\sum\limits_{\tau}}
\left\{  \left(  d_{L\tau\uparrow}^{\dag}d_{R\overline{\tau}\downarrow}^{\dag
}-d_{L\overline{\tau}\downarrow}^{\dag}d_{R\tau\uparrow}^{\dag}\right)
+h.c.\right\}  \label{Hprox}%
\end{equation}
$d_{i\tau\sigma}^{\dag}$ the creation operator for an electron with spin
$\sigma$ in orbital $\tau$ of dot $i\in\{L,R\}$ and $n_{i\tau\sigma}%
=d_{i\tau\sigma}^{\dag}d_{i\tau\sigma}$. For simplicity, one can assume that
the orbital energies in dots $L$ and $R$ are both equal to $\varepsilon$ in
the absence of the external microwave irradiation, which can be obtained by
tuning properly the dots' DC gate voltages $V_{g}^{L(R)}$. The term
$\Delta_{so}$ is caused by spin-orbit coupling inside the carbon
nanotube\cite{Jespersen}. The term $\Delta_{K\leftrightarrow K^{\prime}}$
describes a coupling between the $K$ and $K^{\prime}$ orbitals of dot $i$, due
to disorder at the level of the nanotube atomic structure
\cite{Liang,Kuemmeth,Jespersen,Palyi}. The term in $t_{ee}$ describes interdot
hopping. The term $H_{int}$ accounts for Coulomb charging effects. One can
assume that there cannot be more than one electron in each dot, due to a
strong intra-dot Coulomb charging energy. Therefore, Cooper pairs injected
inside the CPS are split into two electrons, one in each dot. The term
$H_{prox}$ accounts for coherent injection of singlet Cooper pairs inside the
double dot \cite{Eldridge}. This approach is valid provided quasiparticle
transport between the superconducting contact and the double dot can be
disregarded. This requires $eV_{b}<\Delta$, with $\Delta$ the BCS gap of the
superconducting contact. The hamiltonian $H_{DQD}$ must be supplemented by the
normal leads hamiltonian
\begin{equation}
H_{leads}=%
{\textstyle\sum\nolimits_{k_{\tau},\tau,i,\sigma}}
\varepsilon_{ik_{\tau}}c_{ik_{\tau}\sigma}^{\dag}c_{ik_{\tau}\sigma}+h.c.
\end{equation}
and the tunnel coupling between the dots and normal leads
\begin{equation}
H_{t}=%
{\textstyle\sum\nolimits_{k_{\tau},\tau,i,\sigma}}
tc_{ik_{\tau}\sigma}^{\dag}d_{i\tau\sigma}+h.c.
\end{equation}
with $c_{ik_{\tau}\sigma}$ the annihilation operator for an electron with spin
$\sigma$ in orbital $k_{\tau}$ of the normal lead $i\in\{L,R\}$.

The effect of the microwave gate voltage bias can also be described with
hamiltonian terms. The gate voltage $V_{ac}^{L(R)}(t)=v_{ac}^{L(R)}\sin
(\omega_{RF}t)$ corresponds to an electric field $E_{ac}^{L(R)}=V_{ac}%
^{L(R)}(t)/d$, with $d$ the center to ground separation of the waveguide
providing the microwave signal. This also corresponds in the Coulomb gauge to
a vector potential $A_{ac}^{L(R)}=-v_{ac}^{L(R)}\cos(\omega_{RF}t)/\omega
_{RF}d$ on dot $L(R)$, which is assumed to be perpendicular to the carbon
nanotube. The interplay between $A_{ac}^{L(R)}$ and intersubband spin-orbit
coupling elements induced by the nanotube curvature results in a spin/photon
coupling term (see Ref. \cite{Epaps} for details)
\begin{equation}
H_{RF}^{so}=-%
{\displaystyle\sum\limits_{i,\tau,\sigma}}
e\alpha_{i\tau\sigma}v_{ac}^{i}\cos(\omega_{RF}t)d_{i\tau\sigma}^{\dag
}d_{i\tau\overline{\sigma}}\label{hso}%
\end{equation}
with $e>0$ the electron charge. For simplicity, this article uses the
particular structure $\alpha_{i\tau\sigma}=\mathbf{i}\sigma\alpha_{i}$ with
$\alpha_{i}\in\mathbb{R}$ and $\mathbf{i}$ the imaginary unit number, obtained
from a microscopic description of spin-orbit coupling in a zigzag nanotube
quantum dot \cite{Epaps}, based on Refs.\cite{Izumida,Klinovaja} (see also
Refs. \onlinecite{Huertas,Ando,DeMartino,Jeong,Bulaev}). However, part VII.1
will show that the results presented here can be generalized straightforwardly
to a more general $\alpha_{i\tau\sigma}$. The dimensionless coefficient
$\alpha_{i}$ corresponds to the coefficient $\lambda_{i}/eV_{rms}$ of
reference \cite{CKLY}, with $V_{rms}$ the amplitude of vacuum voltage
fluctuations for the photonic cavity considered in this reference. The value
of $\alpha_{i}$ can be estimated to typically $3.10^{-4}$ while $v_{ac}%
^{L(R)}$ can reach typically $100\mathrm{\mu V}$. One can also use a
hamiltonian term $H_{RF}^{g}$ to account for the modulation of the dots
orbital energies by the microwave gate voltages. For simplicity, one can
disregard the mutual capacitive coupling between the two dots. In this case,
one finds
\begin{equation}
H_{RF}^{g}=-%
{\displaystyle\sum\limits_{i,\tau,\sigma}}
\kappa_{i}ev_{ac}^{i}\sin(\omega_{RF}t)n_{i\tau\sigma}%
\end{equation}
where $\kappa_{i}$ is a dimensionless capacitive coupling constant which is
typically of the order of $10^{-2}$.

In the following, it is assumed that electrons can go from dot $L(R)$ to the
corresponding normal metal contact but not the reverse. This can be obtained
by using a bias voltage $V_{b}$ such that
\begin{equation}
eV_{b}>2\Delta_{r}+t_{ee}+\frac{1}{2}\sqrt{8t_{eh}^{2}+(\delta-2\Delta
_{r})^{2}}+\lambda k_{B}T \label{condVb}%
\end{equation}
with $\Delta_{r}=\sqrt{\Delta_{so}^{2}+\Delta_{K\leftrightarrow K^{\prime}%
}^{2}}$ and $\lambda$ a dimensionless coefficient which takes into account the
effective thermal broadening of the levels (see Ref. \cite{Epaps} for details).

\section{III. Expression of the even-charged CPS eigenstates}

This section discusses the relevant eigenstates of $H_{DQD}^{eff}$ in the even
charge sector for $\delta\sim2\Delta_{r}$, with $\delta=2\varepsilon$ the
energy of a CPS doubly occupied state for $t_{eh}=\Delta_{so}=\Delta
_{K\leftrightarrow K^{\prime}}=0$. The parameter $\delta$ can be tuned with
$V_{g}^{L(R)}$. \begin{figure}[h]
\includegraphics[width=1.\linewidth]{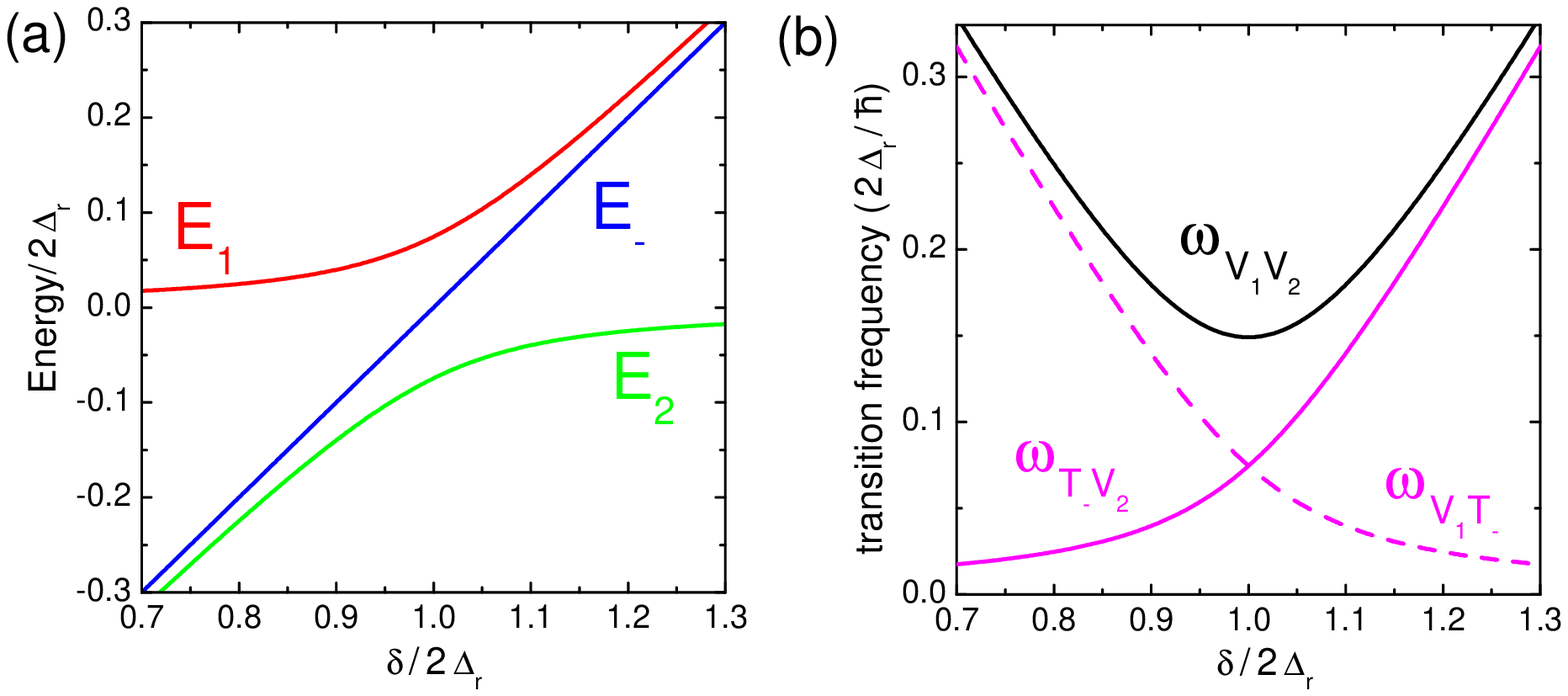}\newline\includegraphics
[width=0.7\linewidth]{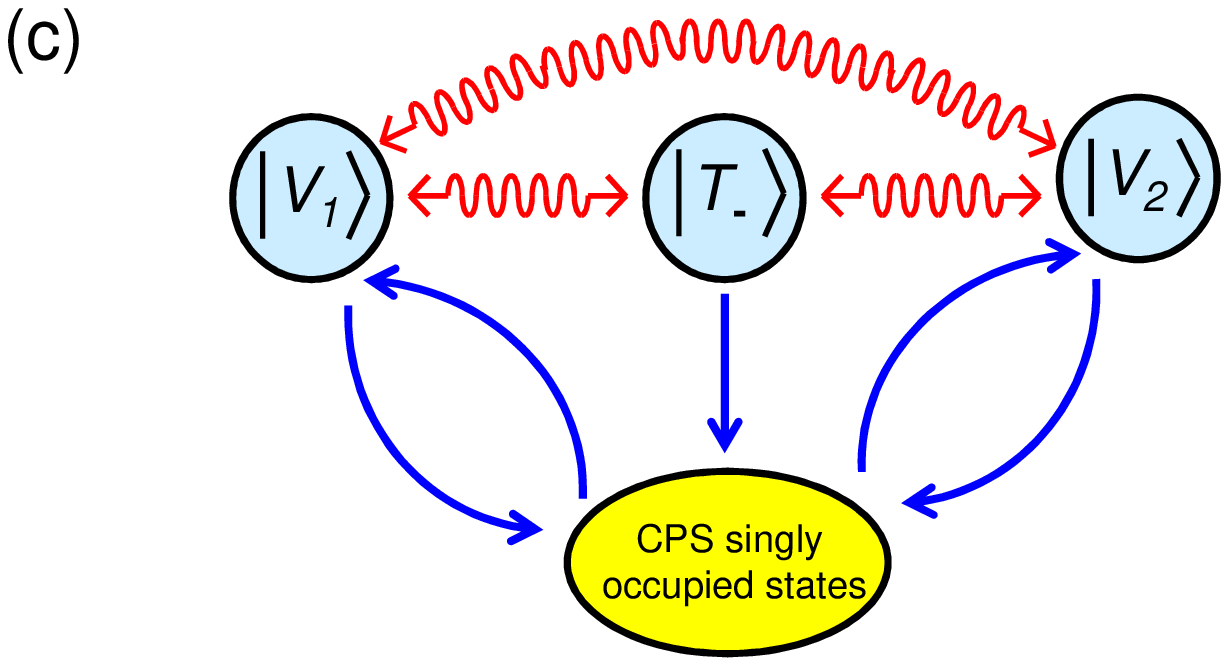}\caption{{}(a) Energies $E_{1}$,
$E_{2}$ and $E_{-}$ of the states $\left\vert V_{1}\right\rangle $,
$\left\vert V_{2}\right\rangle $ and $\left\vert T_{-}\right\rangle $ as a
function of $\delta$. (b) Transition frequencies $\omega_{V_{1}T_{-}}$,
$\omega_{T_{-}V_{2}}$ and $\omega_{V_{1}V_{2}}$ of the CPS as a function of
$\delta$. (c) Dynamics of the CPS near the working point $\delta=2\Delta_{r}$.
We consider a bias voltage regime such that the tunnel transitions between the
different CPS states (blue arrows) occur together with the transfer of one
electron towards the normal contacts. A microwave irradiation can induce
transitions between the states $\left\vert V_{1}\right\rangle $ and
$\left\vert V_{2}\right\rangle $, $\left\vert V_{1}\right\rangle $ and
$\left\vert T_{-}\right\rangle $, or $\left\vert T_{-}\right\rangle $ and
$\left\vert V_{2}\right\rangle $ without any transfer of electrons between the
CPS and the leads (red wavy arrows). We have used $t_{eh}/\Delta_{so}=1/3$ and
$\Delta_{K/K^{\prime}}/\Delta_{so}=6$ in panels (a) and (b).}%
\end{figure}The coupling $t_{eh}$ hybridizes the CPS empty state $\left\vert
0,0\right\rangle $ with the subspace of the CPS doubly occupied states
\{$\left\vert \tau\sigma,\tau^{\prime}\sigma\right\rangle $\}, where
$\left\vert \tau\sigma,\tau^{\prime}\sigma\right\rangle $ denotes a CPS state
with one electron with spin $\sigma$ in orbital $\tau$ of dot L and one
electron with spin $\sigma^{\prime}$ in orbital $\tau^{\prime}$ of dot R
\cite{Recher:11,Godschalk}. The resulting even-charged subspace is called
$\mathcal{\tilde{E}}$. Near the working point $\delta\sim2\Delta_{r}$, the CPS
dynamics involves a subspace $\mathcal{E}$ of at maximum five eigenstates from
$\mathcal{\tilde{E}}$. Three of these eigenstates have an energy $E_{-}%
=\delta-2\Delta_{r}$, namely
\begin{align}
\left\vert T_{0}\right\rangle  &  =\sum_{\sigma}\frac{1}{2}(\sigma\frac
{\Delta_{so}}{\Delta_{r}}-1)\left\vert \mathcal{C}_{+}(K\sigma,K^{\prime}%
\bar{\sigma})\right\rangle \label{T0}\\
&  +\frac{\Delta_{K/K^{\prime}}}{2\Delta_{r}}%
{\displaystyle\sum\limits_{\tau}}
\left\vert \mathcal{C}_{+}(\tau\uparrow,\tau\downarrow)\right\rangle \nonumber
\end{align}%
\begin{align}
\left\vert T_{+}\right\rangle  &  =\sum_{\sigma}\frac{1}{2}\left(
\frac{\Delta_{so}}{\Delta_{r}}-\sigma\right)  \frac{\left\vert K\sigma
,K\sigma\right\rangle -\left\vert K^{\prime}\bar{\sigma},K^{\prime}\bar
{\sigma}\right\rangle }{\sqrt{2}}\label{Tplus}\\
&  +\sum_{\sigma}\sigma\frac{\Delta_{K/K^{\prime}}}{2\Delta_{r}}\left\vert
\mathcal{C}_{+}(K\sigma,K^{\prime}\sigma)\right\rangle \nonumber
\end{align}
and
\begin{align}
\left\vert T_{-}\right\rangle  &  =\sum_{\sigma}\frac{1}{2}\left(
\frac{\Delta_{so}}{\Delta_{r}}\sigma-1\right)  \frac{\left\vert K\sigma
,K\sigma\right\rangle +\left\vert K^{\prime}\bar{\sigma},K^{\prime}\bar
{\sigma}\right\rangle }{\sqrt{2}}\label{Tmoins}\\
&  +\sum_{\sigma}\frac{\Delta_{K/K^{\prime}}}{2\Delta_{r}}\left\vert
\mathcal{C}_{+}(K\sigma,K^{\prime}\sigma)\right\rangle \nonumber
\end{align}
where $\bar{\sigma}$ denotes the spin direction opposite to $\sigma$ and
$\left\vert \mathcal{C}_{\pm}(\tau\sigma,\tau^{\prime}\sigma^{\prime
})\right\rangle =(\left\vert \tau\sigma,\tau^{\prime}\sigma^{\prime
}\right\rangle \pm\left\vert \tau^{\prime}\sigma^{\prime},\tau\sigma
\right\rangle )/\sqrt{2}$. The two remaining eigenstates
\begin{equation}
\left\vert V_{1}\right\rangle =\sqrt{1-v_{1}^{2}}\left\vert 0,0\right\rangle
+v_{1}\left\vert \mathcal{S}\right\rangle
\end{equation}
and
\begin{equation}
\left\vert V_{2}\right\rangle =\sqrt{1-v_{2}^{2}}\left\vert 0,0\right\rangle
+v_{2}\left\vert \mathcal{S}\right\rangle
\end{equation}
have eigenenergies
\begin{equation}
E_{1(2)}=\frac{1}{2}\left(  \delta-2\Delta_{r}\pm\sqrt{8t_{eh}^{2}%
+(\delta-2\Delta_{r})^{2}}\right)  \label{E12}%
\end{equation}
with%
\begin{align}
\left\vert \mathcal{S}\right\rangle  &  =\sum_{\sigma}\left\{  \frac{1}%
{2}(\frac{\Delta_{so}}{\Delta_{r}}-\sigma)\left\vert \mathcal{C}_{-}%
(K\sigma,K^{\prime}\bar{\sigma})\right\rangle \right\} \label{singlet}\\
&  +\frac{\Delta_{K/K^{\prime}}}{2\Delta_{r}}%
{\displaystyle\sum\limits_{\tau}}
\left\vert \mathcal{C}_{-}(\tau\uparrow,\tau\downarrow)\right\rangle \nonumber
\end{align}
and
\begin{equation}
v_{1(2)}=\frac{2t_{eh}}{\sqrt{8t_{eh}^{2}+(\delta-2\Delta_{r})(\delta
-2\Delta_{r}\mp\sqrt{8t_{eh}^{2}+(\delta-2\Delta_{r})^{2}})}} \label{v12}%
\end{equation}
The existence of the $K/K^{\prime}$ degree of freedom complicates slightly the
definition of the CPS eigenstates. However, from the definition of $\left\vert
\mathcal{C}_{\pm}(\tau\sigma,\tau^{\prime}\sigma^{\prime})\right\rangle $, one
can see that $\left\vert \mathcal{S}\right\rangle $ corresponds to a
generalized spin-singlet state whereas $\left\vert T_{0}\right\rangle $,
$\left\vert T_{-}\right\rangle $ and $\left\vert T_{+}\right\rangle $
correspond to generalized spin-triplet states. The coupling $t_{eh}$
hybridizes the empty state $\left\vert 0,0\right\rangle $ with $\left\vert
\mathcal{S}\right\rangle $ only, due to the hypothesis that the
superconducting contact injects spin-singlet pairs inside the CPS. Figure 2.a
shows the energies $E_{1}$, $E_{2}$ and $E_{-}$ as a function of $\delta$. The
energies $E_{1}$ and $E_{2}$ show an anticrossing with a width $2\sqrt
{2}t_{eh}$ at $\delta=2\Delta_{r}$, due to the coherent coupling between
$\left\vert 0,0\right\rangle $ and $\left\vert \mathcal{S}\right\rangle $. The
energy $E_{-}$ of the triplet states lies between $E_{1}$ and $E_{2}$. Figure
2.b shows the transition frequencies $\omega_{V_{1}T_{-}}$, $\omega
_{T_{-}V_{2}}$ and $\omega_{V_{1}V_{2}}$ of the CPS, with $\omega_{m^{\prime
}m}=(E_{m^{\prime}}-E_{m})/\hbar$. These frequencies will play an important
role in the following.

\section{IV. Microwave-induced matrix elements}

This section discusses the effect of the microwave gate bias on the
eigenstates defined in section III. Inside the subspace $\mathcal{E}$,
$H_{RF}^{so}$ has only three finite matrix elements, i.e.
\begin{align}
&  \left\langle T_{_{-}}\right\vert H_{RF}^{so}\left\vert V_{1(2)}%
\right\rangle \label{alpha1}\\
&  =-\mathbf{i}ev_{1(2)}\frac{\Delta_{K\leftrightarrow K^{\prime}}}{\Delta
_{r}}(\alpha_{L}v_{ac}^{L}-\alpha_{R}v_{ac}^{R})\cos(\omega_{RF}t)\nonumber
\end{align}
and
\begin{equation}
\left\langle T_{_{+}}\right\vert H_{RF}^{so}\left\vert T_{0}\right\rangle
=\mathbf{i}e\frac{\Delta_{K\leftrightarrow K^{\prime}}}{\Delta_{r}}(\alpha
_{L}v_{ac}^{L}+\alpha_{R}v_{ac}^{R})\cos(\omega_{RF}t)\label{alpha2}%
\end{equation}
These terms are finite because $H_{RF}^{so}$ flips the spins in the dots. The
minus sign in Eq.(\ref{alpha1}) is a direct consequence of the fact that
$\left\vert V_{1(2)}\right\rangle $ comprises a singlet component whereas
$\left\vert T_{_{-}}\right\rangle $ is a triplet state. In contrast, the plus
sign in Eq.(\ref{alpha2}) is due to the fact that $\left\vert T_{0}%
\right\rangle $ and $\left\vert T_{_{+}}\right\rangle $ are both triplet
states. The matrix element of Eq.(\ref{alpha2}) is always non-resonant since
it couples two states with the same energy. Therefore, it can be disregarded
in our study. The hamiltonian $H_{RF}^{g}$ has only one finite coupling
element in the subspace $\mathcal{E}$, i.e.
\begin{equation}
\left\langle V_{1}\right\vert H_{RF}^{g}\left\vert V_{2}\right\rangle
=-v_{1}v_{2}e(\kappa_{L}v_{ac}^{L}+\kappa_{R}v_{ac}^{R})\sin(\omega
_{RF}t)\label{8}%
\end{equation}
with $v_{1}v_{2}=\sqrt{2}t_{eh}/\sqrt{8t_{eh}^{2}+(\delta-2\Delta_{r})^{2}}$.
The addition of $\kappa_{L}v_{ac}^{L}$ and $\kappa_{R}v_{ac}^{R}$ in Eq.
(\ref{8}) is due to the fact that the double occupation energy $\delta$ is
shifted by $-(\kappa_{L}V_{ac}^{L}(t)+\kappa_{R}V_{ac}^{R}(t))$ when a
microwave excitation is applied to the device.

One can find experimental means to have $V_{ac}^{L}$ and $V_{ac}^{R}$ on
phase, in agreement with the assumption made in section II. In this case, the
matrix element $\left\langle T_{_{-}}\right\vert H_{RF}^{so}\left\vert
V_{1(2)}\right\rangle $ vanishes when $\alpha_{L}v_{ac}^{L}=\alpha_{R}%
v_{ac}^{R}$. This effect is directly related to the injection of coherent
singlet Cooper pairs inside the CPS since it is due to the existence of the
minus sign in Eq. (\ref{alpha1}). If the injected pairs were in a product
state instead of an entangled state, the matrix element (\ref{alpha1}) would
not be subradiant (see section VII.4). Therefore, coherent pair injection
inside the CPS can be revealed by observing microwave-induced transitions
between $\left\vert V_{1(2)}\right\rangle $ and $\left\vert T_{_{-}%
}\right\rangle $, and checking that these transitions are suppressed for
$\alpha_{L}v_{ac}^{L}=\alpha_{R}v_{ac}^{R}$. The following sections describe
how to probe these microwave-induced transitions with a DC current measurement.

\section{V. Master equation description of the CPS dynamics}

In the following, the states $\left\vert T_{0}\right\rangle $ and $\left\vert
T_{+}\right\rangle $ are disregarded because they are not populated in simple
limits where relaxation towards them is neglected. The sequential tunneling
limit $\Gamma_{N}\ll k_{B}T$ is furthermore assumed, with $\Gamma_{N}$ the
tunnel escape rate of an electron from one of the dots to the corresponding
normal lead. For simplicity, it is assumed that this rate does not depend on
the dot orbital and spin indices. This would change only quantitatively the
results shown in this paper. In the absence of microwave irradiation, the
dynamics of the CPS can be described with a master
equation\cite{Eldridge,Sauret:2004}
\begin{equation}
\frac{dP}{dt}=MP
\end{equation}
with
\begin{equation}
P=\left[
\begin{array}
[c]{c}%
P_{V_{1}}\\
P_{V_{2}}\\
P_{T_{-}}\\
P_{\text{single}}%
\end{array}
\right]
\end{equation}
and
\begin{equation}
M=\left[
\begin{array}
[c]{cccc}%
-2v_{1}^{2}\Gamma_{N} & 0 & 0 & (1-v_{1}^{2})\Gamma_{N}\\
0 & -2v_{2}^{2}\Gamma_{N} & 0 & (1-v_{2}^{2})\Gamma_{N}\\
0 & 0 & -2\Gamma_{N} & 0\\
2v_{1}^{2}\Gamma_{N} & 2v_{2}^{2}\Gamma_{N} & 2\Gamma_{N} & -\Gamma_{N}%
\end{array}
\right]  \label{M}%
\end{equation}
Above, $P_{i}$ denotes the probability of state $\left\vert i\right\rangle $,
with $i\in\{V_{1},V_{2},T_{-}\}$. The vector $P$ also includes the global
probability $P_{\text{single}}$ of having a double dot singly occupied state.
The use of this global probability is sufficient to describe the dynamics of
the CPS because the single electron tunnel rate $\Gamma_{N}$ to the normal
leads is assumed to be independent from the dot orbital and spin indices. The
various singly occupied eigenstates of $H_{DQD}^{eff}$ are defined in section
VII.3. The exact relation $v_{1}^{2}+v_{2}^{2}=1$ has been used to simplify
the above expression of $M$.

The microwave excitation $H_{RF}^{so}$ can induce resonances between the
states $\left\vert V_{1(2)}\right\rangle $ and $\left\vert T_{-}\right\rangle
$ while the excitation $H_{RF}^{g}$ couples $\left\vert V_{1}\right\rangle $
and $\left\vert V_{2}\right\rangle $. One can use a rotating frame
approximation on independent resonances to describe these effects. This
approach is valid provided one of the microwave-induced resonance has a
dominant effect on the others, which requires the frequencies $\omega
_{V_{1}T_{-}}$, $\omega_{T_{-}V_{2}}$ and $\omega_{V_{1}V_{2}}$ to be
sufficiently different. The rotating frame approximation also requires to use
small amplitudes $\kappa_{L(R)}v_{ac}^{L(R)}$ and $\alpha_{L(R)}v_{ac}^{L(R)}$
compared to $\omega_{V_{1}T_{-}}$, $\omega_{T_{-}V_{2}}$, $\omega_{V_{1}V_{2}%
}$ and $\omega_{RF}$. In this case, the stationary state occupation
probabilities can be obtained from
\begin{equation}
0=(M+M_{RF})P_{stat}%
\end{equation}
with
\begin{align}
&  M_{RF}\\
&  =\left[
\begin{array}
[c]{cccc}%
-r_{V_{1}T_{-}}-r_{V_{1}V_{2}} & r_{V_{1}V_{2}} & r_{V_{1}T_{-}} & 0\\
r_{V_{1}V_{2}} & -r_{T_{-}V_{2}}-r_{V_{1}V_{2}} & r_{T_{-}V_{2}} & 0\\
r_{V_{1}T_{-}} & r_{T_{-}V_{2}} & -r_{V_{1}T_{-}}-r_{T_{-}V_{2}} & 0\\
0 & 0 & 0 & 0
\end{array}
\right]
\end{align}%
\begin{equation}
r_{ab}(\omega)=\frac{\left\vert C_{ab}\right\vert ^{2}}{\hbar^{2}}%
\frac{2\Gamma_{ab}}{(\omega-\omega_{ab})^{2}+\Gamma_{ab}^{2}}>0
\end{equation}%
\begin{align}
C_{V_{1}T_{_{-}}} &  =v_{1}e\frac{\Delta_{K\leftrightarrow K^{\prime}}%
}{2\Delta_{r}}(\alpha_{L}v_{ac}^{L}-\alpha_{R}v_{ac}^{R})\\
C_{T_{_{-}}V_{2}} &  =v_{2}e\frac{\Delta_{K\leftrightarrow K^{\prime}}%
}{2\Delta_{r}}(\alpha_{L}v_{ac}^{L}-\alpha_{R}v_{ac}^{R})\\
C_{V_{1}V_{2}} &  =\frac{v_{1}v_{2}}{2}e(\kappa_{L}v_{ac}^{L}+\kappa_{R}%
v_{ac}^{R})
\end{align}
and $%
{\textstyle\sum\nolimits_{i}}
P_{stat,i}=1$. Above, $\Gamma_{ab}$ corresponds to the coherence time between
the states $\left\vert a\right\rangle $ and $\left\vert b\right\rangle $.
Assuming that $\Gamma_{ab}$ is limited by tunneling to the normal leads,
one obtains $\Gamma_{V_{1}V_{2}}=\Gamma_{N}$, $\Gamma_{V_{1}T_{-}}%
=(1+v_{1}^{2})\Gamma_{N}$ and $\Gamma_{T_{-}V_{2}}=(1+v_{2}^{2})\Gamma_{N}$.

Figure 2.c represents schematically the dynamics of the CPS near the working
point $\delta=2\Delta_{r}$. Due to the assumptions made in section II on
$V_{b}$, the tunnel transitions between the different CPS states (blue arrows)
always occur together with the transfer of one electron towards one of the
normal metal contacts. In contrast, the microwave irradiation induces
transitions between the states $\left\vert V_{1}\right\rangle $ and
$\left\vert V_{2}\right\rangle $, $\left\vert V_{1}\right\rangle $ and
$\left\vert T_{-}\right\rangle $, or $\left\vert T_{-}\right\rangle $ and
$\left\vert V_{2}\right\rangle $ without any exchange of electrons with the
normal contacts (red wavy arrows). The state $\left\vert T_{-}\right\rangle $
can be reached through a microwave-induced transition but not through a tunnel
process because it has no component in $\left\vert 0,0\right\rangle $. The
states $\left\vert V_{1}\right\rangle $ and $\left\vert V_{2}\right\rangle $
can be both reached or left through a tunnel event because they have
components in both $\left\vert 0,0\right\rangle $ and $\left\vert
\mathcal{S}\right\rangle $.

\section{VI. Results}

\subsection{VI.1 Principle of the measurement}

From Eq. (\ref{M}), the tunnel rate transitions from the states $\left|
V_{1}\right\rangle $, $\left|  V_{2}\right\rangle $, and $\left|
T_{-}\right\rangle $ to the ensemble of the singly occupied states are
$2v_{1}^{2}\Gamma_{N}$, $2v_{2}^{2}\Gamma_{N}$ and $2\Gamma_{N}$ respectively,
while the tunnel transition rate from a singly occupied state to $\left|
V_{1}\right\rangle $ or $\left|  V_{2}\right\rangle $ is $\Gamma_{N}$. As a
result, the DC current $I$ flowing at the input of the CPS can be calculated
as
\begin{equation}
I=R.P_{stat} \label{cur}%
\end{equation}
with $R=e\Gamma_{N}[2v_{1}^{2},2v_{2}^{2},2,1]$. Figure 3.a shows the
coefficients $v_{1}^{2}$ and $v_{2}^{2}$ as a function of $\delta$. One can
conclude from this plot that except at $\delta=2\Delta_{r}$, the various
components of $R$ have different values. Therefore, a microwave excitation
changing the population of the states $\left|  V_{1}\right\rangle $, $\left|
V_{2}\right\rangle $, and $\left|  T_{-}\right\rangle $ should affect the
value of the DC current flowing through the CPS. This effect will be used in
the following to reveal the microwave-induced transitions between $\left|
V_{1}\right\rangle $, $\left|  V_{2}\right\rangle $, and $\left|
T_{-}\right\rangle $.

\subsection{VI.2 Stationary CPS state occupations}

Let us first discuss the dependence of the CPS state probabilities $P_{i}$ on
the parameter $\delta$ for $V_{ac}^{L(R)}=0$ (see Fig.3, black dotted lines in
the three lowest panels). To understand this dependence, one must keep in mind
the fact that the tunnel rate transitions from the states $\left|
V_{1}\right\rangle $ and $\left|  V_{2}\right\rangle $ to the ensemble of the
singly occupied states are $2v_{1}^{2}\Gamma_{N}$ and $2v_{2}^{2}\Gamma_{N}$,
as already discussed in section VI.1. For $\delta$ well below $2\Delta_{r}$,
$v_{1}^{2}$ tends to zero. As a result, the CPS cannot escape easily from the
state $\left|  V_{1}\right\rangle $, whose probability tends to 1. This is
because in this limit, the state $\left|  V_{1}\right\rangle $ is almost equal
to the empty state $\left|  0,0\right\rangle $, which makes the emission of an
electron towards the normal leads very difficult. On opposite, for $\delta$
well above $2\Delta_{r}$, it is the probability of the state $\left|
V_{2}\right\rangle $ which tends to one because $\left|  V_{2}\right\rangle $
tends to $\left|  0,0\right\rangle $. In the absence of a microwave
irradiation, the probability of state $\left|  T_{-}\right\rangle $ remains
equal to zero since transitions towards these state are not possible.
\begin{figure}[h]
\includegraphics[width=1.0\linewidth]{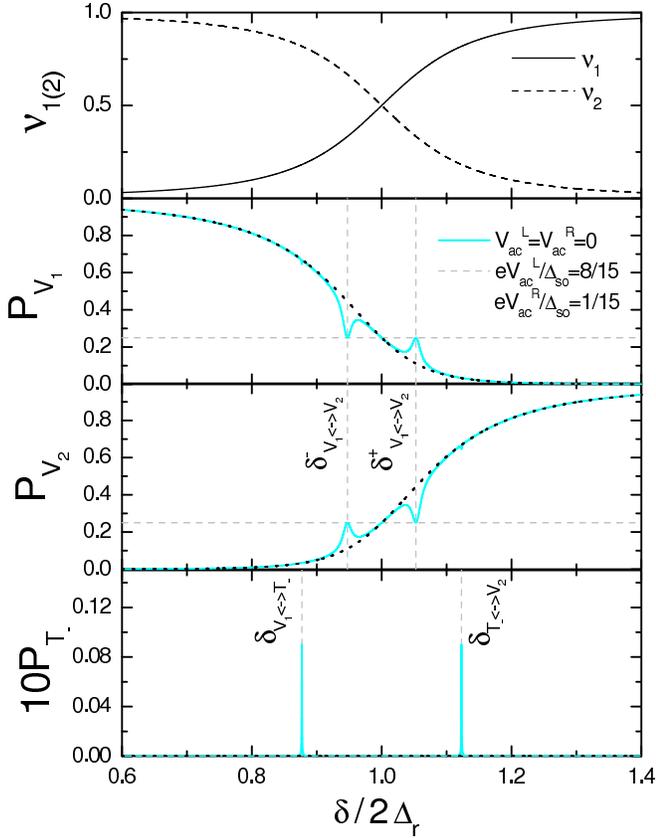}\caption{{}Coefficients
$v_{1(2)}^{2}$, and probabilities $P_{V_{1}}$, $P_{V_{2}}$ and $P_{T_{-}}$ of
the CPS states $\left|  V_{1}\right\rangle $, $\left|  V_{2}\right\rangle $
and $\left|  T_{-}\right\rangle $ as a function of $\delta$. We have used
$t_{eh}/\Delta_{so}=1/3$, $\Delta_{K/K^{\prime}}/\Delta_{so}=3$, $2\pi
\hbar\Gamma_{N}/\Delta_{so}=1.37~10^{-3}$, $ev_{ac}^{L}/\Delta_{so}=1/15$,
$ev_{ac}^{R}/\Delta_{so}=8/15$, $\alpha_{L}=\alpha_{R}=3.10^{-4}$, $\kappa
_{L}=\kappa_{R}=10^{-2}$ and $\omega_{RF}=3t_{eh}$.}%
\end{figure}

Let us now discuss the case $v_{ac}^{L(R)}$ finite (see Fig. 3, red full lines
in the three lowest panels). The term $H_{RF}^{g}$ excites the $\left\vert
V_{1}\right\rangle \leftrightarrow\left\vert V_{2}\right\rangle $ transition,
which causes peaks or dips in $P_{V_{1}}$ and $P_{V_{2}}$ for $\omega
_{RF}=\omega_{V_{1}V_{2}}$, i.e. $\delta=\delta_{V_{1}\leftrightarrow V_{2}%
}^{\pm}$ with
\begin{equation}
\delta_{V_{1}\leftrightarrow V_{2}}^{\pm}=2\Delta_{r}\pm\sqrt{\omega_{RF}%
^{2}-8t_{eh}^{2}}%
\end{equation}
The term $H_{RF}^{so}$ excites the $\left\vert V_{1}\right\rangle
\leftrightarrow\left\vert T_{-}\right\rangle $ \ and $\left\vert
T_{-}\right\rangle \leftrightarrow\left\vert V_{2}\right\rangle $ transitions,
which causes peaks in $P_{T_{-}}$ for $\omega_{RF}=\omega_{V_{1}T_{-}}$ and
$\omega_{RF}=\omega_{T_{-}V_{2}}$, i.e. $\delta=\delta_{V_{1}\leftrightarrow
T_{-}}$ and $\delta=\delta_{T_{-}\leftrightarrow V_{2}}$ respectively, with
\begin{equation}
\delta_{V_{1}\leftrightarrow T_{-}}=2\Delta_{r}-\omega_{RF}+(2t_{eh}%
^{2}/\omega_{RF})
\end{equation}
and
\begin{equation}
\delta_{T_{-}\leftrightarrow V_{2}}=2\Delta_{r}+\omega_{RF}-(2t_{eh}%
^{2}/\omega_{RF})
\end{equation}
The term $H_{RF}^{so}$ also causes peaks or dips in $P_{V_{1}}$ and $P_{V_{2}%
}$, but they are hardly visible due to the scale used in Fig. 3. The
decoherence rates $\Gamma_{V_{1}T_{-}}$, $\Gamma_{T_{-}V_{2}}$ and
$\Gamma_{V_{1}V_{2}}$ have similar order of magnitudes (between $\Gamma_{N}$
and $2\Gamma_{N}$). However, the width of the peaks or dips caused by
$H_{RF}^{g}$ seems much larger than the width of the peaks caused by
$H_{RF}^{so}$. This is due to the limit $\alpha_{L(R)}\ll\kappa_{L(R)}$
considered here. As long as the different types of resonances are well
separated in frequency, the resonance $\left\vert V_{1}\right\rangle
\leftrightarrow\left\vert V_{2}\right\rangle $ gives probabilities $P_{V_{1}}$
and $P_{V_{2}}$ which tend to the value 1/4 for $r_{V_{1}V_{2}}$ sufficiently
large. In principle, the $\left\vert V_{1}\right\rangle \leftrightarrow
\left\vert T_{-}\right\rangle $ and $\left\vert T_{-}\right\rangle
\leftrightarrow\left\vert V_{2}\right\rangle $ resonances give state
probabilities $P_{i}$ which saturate at more complicated values which depend
on $v_{1(2)}^{2}$ when $r_{V_{1}T_{-}}$ and $r_{T_{-}V_{2}}$ become
sufficiently large. In the regime $\alpha_{L(R)}\ll\kappa_{L(R)}$ considered
here, the $\left\vert V_{1}\right\rangle \leftrightarrow\left\vert
V_{2}\right\rangle $ resonance is saturated while the $\left\vert
V_{1}\right\rangle \leftrightarrow\left\vert T_{-}\right\rangle $ and
$\left\vert T_{-}\right\rangle \leftrightarrow\left\vert V_{2}\right\rangle $
resonances are only weakly excited. This explains that the width of the peaks
or dips related to the $\left\vert V_{1}\right\rangle \leftrightarrow
\left\vert V_{2}\right\rangle $ resonance are much larger.

\subsection{VI.3 Average current at the input of the CPS}

It is useful to discuss first the value $I_{0}$ of the current $I$ at the
input of the CPS in the absence of the microwave excitations. The current
$I_{0}$ can be obtained from Eq.(\ref{cur}) with $v_{ac}^{L(R)}=0$. From Fig.
4.a, $I_{0}$ shows a maximum for $\delta=2\Delta_{r}$, where the two states
$\left\vert V_{1}\right\rangle $ and $\left\vert V_{2}\right\rangle $ both
correspond to equally weighted superpositions of $\left\vert 0,0\right\rangle
$ and $\left\vert \mathcal{S}\right\rangle $. For $\delta$ well below or well
above $2\Delta_{r}$, the current $I_{0}$ vanishes because the CPS is blocked
in the states $\left\vert V_{1}\right\rangle $ or $\left\vert V_{2}%
\right\rangle $, respectively (see section V).

Figure 4.b shows the difference between the current $I$ for a finite microwave
irradiation and $I_{0}$, as a function of $\omega_{RF}$ and $\delta$. The
$\left\vert V_{1}\right\rangle \leftrightarrow\left\vert V_{2}\right\rangle $
transitions yield a broad resonance along the curve $\omega_{RF}=\omega
_{V_{1}V_{2}}=\sqrt{8t_{eh}^{2}+(\delta-2\Delta_{r})^{2}}/\hbar$, which has a
frequency minimum $\omega_{RF}=2\sqrt{2}t_{eh}/\hbar$ at $\delta-2\Delta_{r}$.
However, this resonance vanishes close to $\delta=2\Delta_{r}$ because at this
point, the tunnel escape rates $2v_{1}^{2}\Gamma_{N}$ and $2v_{2}^{2}%
\Gamma_{N}$ of the CPS from $\left\vert V_{1}\right\rangle $ and $\left\vert
V_{2}\right\rangle $ are equal since $v_{1}=v_{2}$ and therefore, the
microwave-induced transitions between the states $\left\vert V_{1}%
\right\rangle $ and $\left\vert V_{2}\right\rangle $ cannot be seen anymore
through a measurement of $I$. The $\left\vert V_{1}\right\rangle
\leftrightarrow\left\vert T_{-}\right\rangle $ and $\left\vert T_{-}%
\right\rangle \leftrightarrow\left\vert V_{2}\right\rangle $ resonances yield
two thinner resonances which cross at the point $O$ corresponding to
$\delta=2\Delta$ and $\hbar\omega_{RF}=\sqrt{2}t_{eh}$. For $\omega_{RF}$
tending to zero, the $\left\vert V_{1}\right\rangle \leftrightarrow\left\vert
T_{-}\right\rangle $ ( $\left\vert T_{-}\right\rangle \leftrightarrow
\left\vert V_{2}\right\rangle $) resonance progressively vanishes from
$I\ $because this corresponds to a regime where the state $\left\vert
V_{1}\right\rangle $ ($\left\vert V_{2}\right\rangle $) is not populated
anymore. Note that the calculation of the current $I$ very close to the point
$O$ is in principle not valid using the rotating wave approximation on
independent resonances since $\omega_{V_{1}T_{-}}=\omega_{T_{-}V_{2}}$ at this
point. However, this represents only an extremely small area of Fig. 4.a (of
order $\Gamma_{N}\times\Gamma_{N}$). Discussing the behavior of the CPS near
point $O$ goes beyond the scope of this paper.

\begin{center}
\begin{figure}[h]
\includegraphics[width=0.5\linewidth]{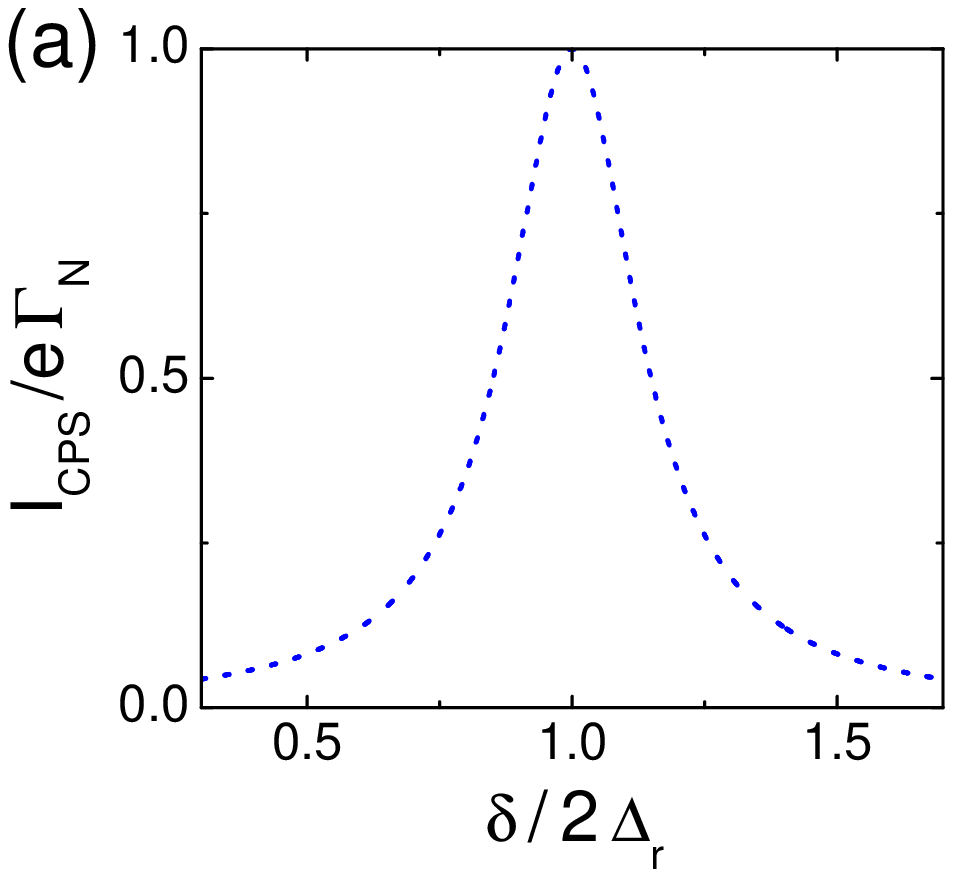}\newline%
\includegraphics[width=1.0\linewidth]{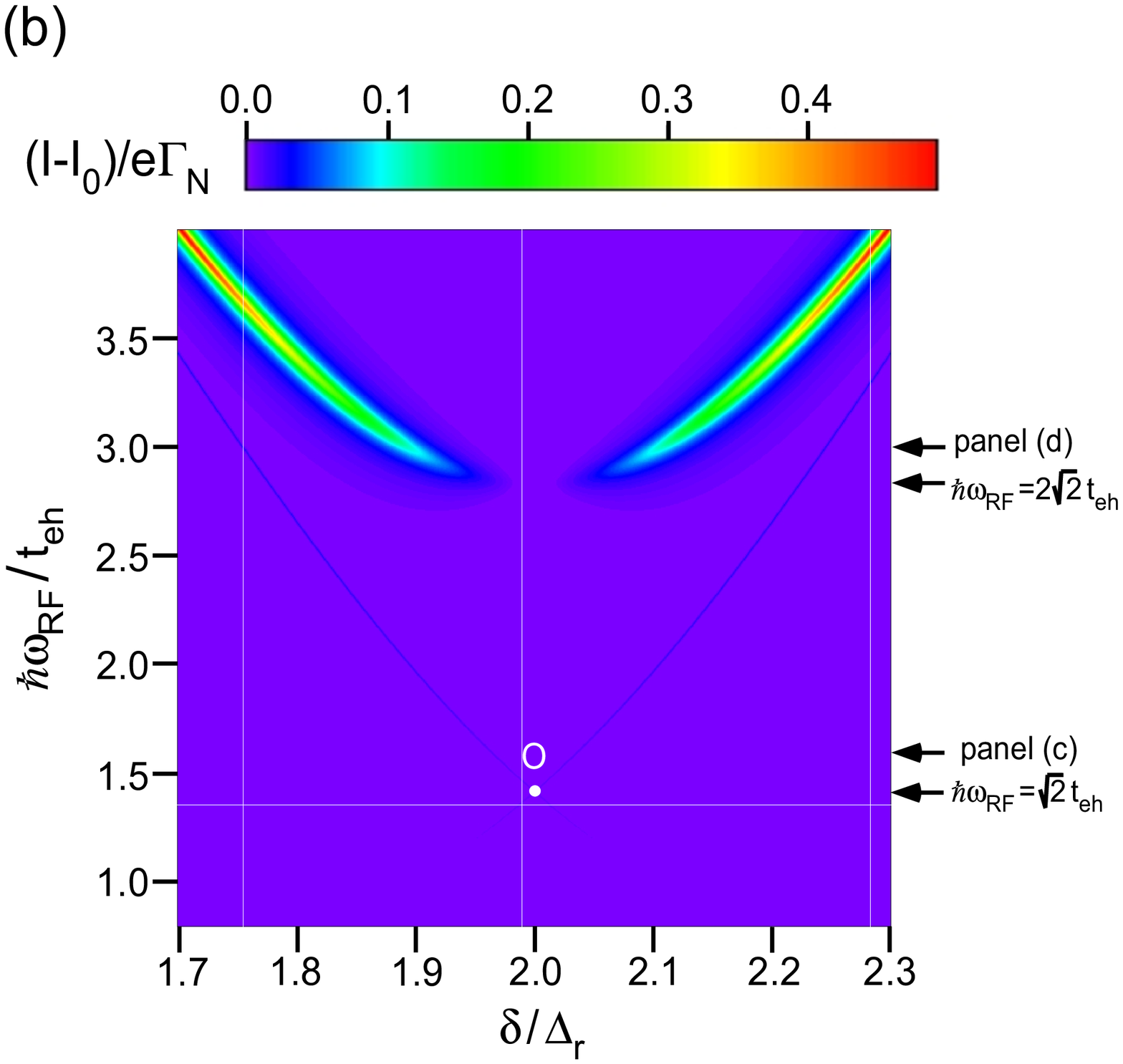}\newline%
\includegraphics
[width=1.\linewidth]{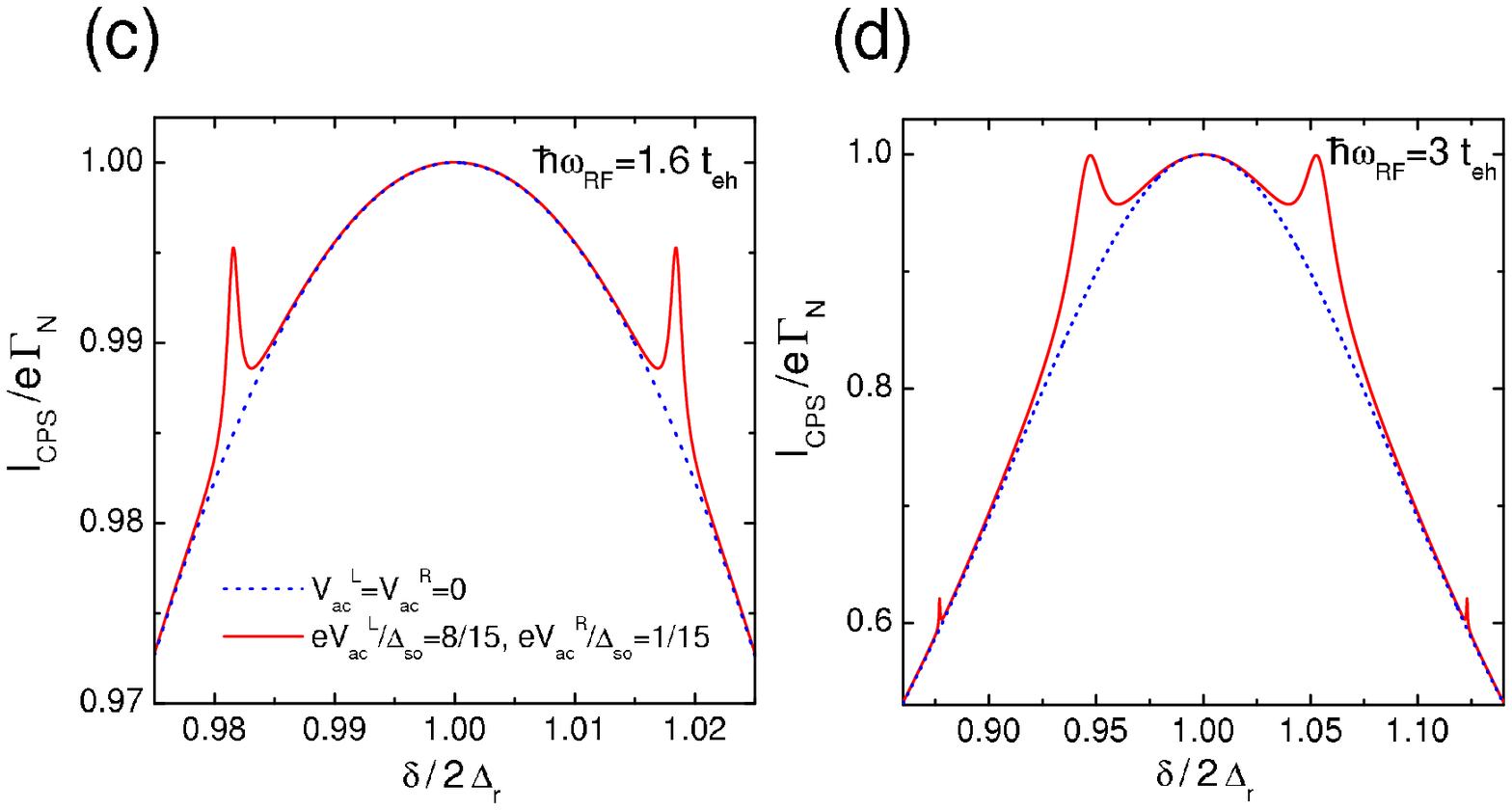}\newline\caption{{}(a): Current $I_{0}$
in the absence of any microwave irradiation as a function of $\delta$ (b):
Difference between the current $I$ for a finite microwave irradiation and the
current $I_{0}$, as a function of $\omega_{RF}$ and $\delta$ (c) and (d):
Current difference $I-I_{0}$ as a function of $\delta$ for $\hbar\omega
_{RF}=1.6t_{eh}$ and $\hbar\omega_{RF}=3t_{eh}$. The other parameters used are
the same as in Fig. 2.}%
\end{figure}
\end{center}

Figures. 4.c and 4.d show $I-I_{0}$ as a function of $\delta$ for two
different values of $\omega_{RF}$. In Fig. 4.c, only the $\left\vert
V_{1}\right\rangle \leftrightarrow\left\vert T_{-}\right\rangle $ and
$\left\vert T_{-}\right\rangle \leftrightarrow\left\vert V_{2}\right\rangle $
resonances are visible because $\omega_{RF}<2\sqrt{2}t_{eh}/\hbar$. In Fig.
4.d, the $\left\vert V_{1}\right\rangle \leftrightarrow\left\vert
V_{2}\right\rangle $ resonances are also visible. The $\left\vert
V_{1}\right\rangle \leftrightarrow\left\vert T_{-}\right\rangle $ and
$\left\vert T_{-}\right\rangle \leftrightarrow\left\vert V_{2}\right\rangle $
resonances appear as much thinner an smaller peaks. At the $\left\vert
V_{1}\right\rangle \leftrightarrow\left\vert V_{2}\right\rangle $ resonances,
for the parameters used in Fig.4.d, $I$ reaches the saturation value
$\Gamma_{N}$ expected for $r_{V_{1}V_{2}}$ large and well separated
resonances. This value can be obtained from Eq.(\ref{cur}), using
$P_{stat}=^{t}[1/4,1/4,0,1/2]$.

\subsection{VI.4 Dependence of the CPS input current on the amplitude of the
microwave irradiation}

This section discusses how the minus sign in Equation (\ref{alpha1}) can be
seen experimentally. One can note $\Delta I_{V_{1}\leftrightarrow V_{2}}^{\pm
}=I(\delta=\delta_{V_{1}\leftrightarrow V_{2}}^{\pm})-I_{0}$, $\Delta
I_{V_{1}\leftrightarrow T_{-}}=I(\delta=\delta_{V_{1}\leftrightarrow T_{-}%
})-I_{0}$ and $\Delta I_{T_{-}\leftrightarrow V_{2}}=I(\delta=\delta
_{T_{-}\leftrightarrow V_{2}})-I_{0}$ the amplitudes of the microwave-induced
current peaks appearing for $\omega_{RF}=\omega_{V_{1}V_{2}}$, $\omega
_{RF}=\omega_{V_{1}T_{-}}$, and $\omega_{RF}=\omega_{T_{-}V_{2}}$. Due to the
symmetries of our model around the point $\delta=2\Delta_{r}$, one has $\Delta
I_{V_{1}\leftrightarrow V_{2}}^{\pm}=\Delta I_{V_{1}\leftrightarrow V_{2}}$
and $\Delta I_{V_{1}\leftrightarrow T_{-}}=\Delta I_{T_{-}\leftrightarrow
_{V_{2}}}$. The top and bottom panels of Fig. 5 show the variations of $\Delta
I_{V_{1}\leftrightarrow T_{-}}$ and $\Delta I_{V_{1}\leftrightarrow V_{2}}$
with $v_{ac}^{R}$ for a constant value of $v_{ac}^{L}$. Due to the plus sign
in Eq. (\ref{8}), $\Delta I_{V_{1}\leftrightarrow V_{2}}$ increases
monotonically with $v_{ac}^{L}$. In Fig.5, this variation is very small
because the $\left\vert V_{1}\right\rangle \leftrightarrow\left\vert
V_{2}\right\rangle $ resonance is already saturated at $v_{ac}^{R}=0$ due to
the value used for $v_{ac}^{L}$. In contrast, due to the minus sign in Eq.
(\ref{alpha1}), $\Delta I_{V_{1}\leftrightarrow T_{-}}$ shows a minimum for
$v_{ac}^{L}=v_{ac}^{R}$. Note that if the electrons pairs injected in the CPS
were not in an entangled state but in a \ product state, such a non-monotonic
behavior would not be possible. For the parameters considered in Fig.5, top
panel, $\Delta I_{V_{1}\leftrightarrow T_{-}}$ vanishes at $v_{ac}^{L}%
=v_{ac}^{R}$ because the effects of the $\left\vert V_{1}\right\rangle
\leftrightarrow\left\vert V_{2}\right\rangle $ resonance can be disregarded.
This should not be true anymore in the case where the different types of
resonances are not well separated, which can happen e.g. if $t_{eh}$ is too
small with respect to the width of the resonances. However, in this case, one
can still expect $\Delta I_{V_{1}\leftrightarrow T_{-}}$ to show a strongly
non-monotonic behavior with a minimum at $v_{ac}^{L}=v_{ac}^{R}$, provided the
couplings $\alpha_{L(R)}$ are sufficiently strong. Treating this case requires
to go beyond the rotating frame approximation with independent resonances used
in this work.

\begin{figure}[h]
\includegraphics[width=0.8\linewidth]{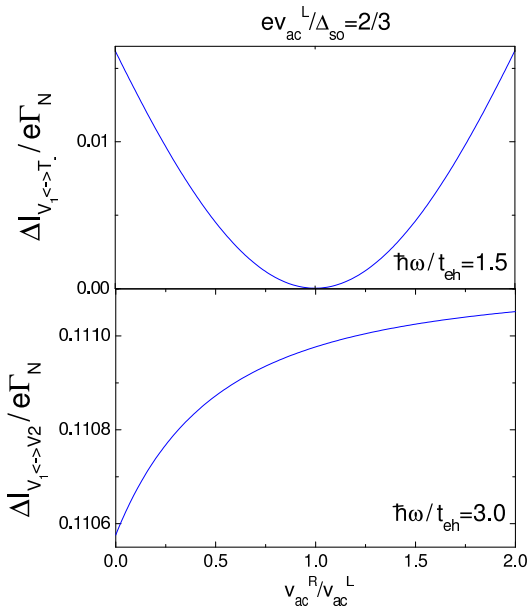}\caption{{}Amplitude of
the current peaks $\Delta I_{V_{1}\leftrightarrow T_{-}}$ \ (top panel) and
$\Delta I_{V_{1}\leftrightarrow V_{2}}$ (bottom panel) as a function of
$v_{ac}^{R}$ for a constant value of $v_{ac}^{L}$, i.e. $ev_{ac}^{L}%
/\Delta_{so}=2/3$. The other parameters used are the same as in Fig. 2.}%
\end{figure}

\subsection{VI.5 Experimental parameters}

This section discusses the parameters used in the Figs. and the order of
magnitude of the signals which can be expected in practice. In Figs. 3 to 5,
the ratio of parameters used correspond for instance to realistic values
$t_{eh}=50~\mathrm{\mu eV}$ (see Refs.
\cite{Hofstetter:09,Hofstetter:11,Schindele,Herrmann:10,Herrmann:12}),
$\Delta_{so}=0.15~\mathrm{meV}$, $\Delta_{K/K^{\prime}}=0.45~\mathrm{meV}$ and
$\Gamma_{N}=50$ \textrm{M}$\mathrm{Hz}$ (see Refs.
\cite{Liang,Kuemmeth,Jespersen}). Note that $\Gamma_{N}=50$ \textrm{M}%
$\mathrm{Hz}$ corresponds to $6.5$ \textrm{mK}, therefore the sequential
tunneling approximation used in this work is valid using for instance $T=65$
\textrm{mK}. In this case, using $\lambda=5$, the condition (\ref{condVb}) to
have electrons flowing only from the dot to the leads and not the reverse
gives $V_{b}>1.05~\mathrm{mV}$ (see section II). This is compatible with the
condition $V_{b}<\Delta$ for having no quasiparticle transport between the
superconducting lead and the dots, by using for instance a Nb contact for
which $\Delta\simeq1.4~\mathrm{meV}$ or a NbN contact for which $\Delta
\simeq3~\mathrm{meV}$. Using the above parameters, the ratio $v_{ac}%
^{L(R)}/\Delta_{so}$ used in Figs. 3 and 4 corresponds to realistic microwave
amplitudes $v_{ac}^{L}=10\mathrm{\mu V}$ and $v_{ac}^{R}=80\mathrm{\mu V}$.
Besides, the maximum frequency $\omega_{RF}=4t_{eh}$ considered in this work
(see Fig. 4.b) corresponds to $48.5~\mathrm{GHz}$, and the frequency at point
O corresponds to$\ 17~\mathrm{GHz}$, which is accessible with current
microwave technologies\cite{Meyer}. Using the above parameters, the amplitude
of the current peaks $\Delta I_{V_{1}\leftrightarrow V_{2}}$ and $\Delta
I_{V_{1}\leftrightarrow T_{-}}$ shown in Fig. 3.d are $\Delta I_{V_{1}%
\leftrightarrow V_{2}}=884~\mathrm{fA}$ and $\Delta I_{V_{1}\leftrightarrow
T_{-}}=207~\mathrm{fA}$ over a background $I_{0}$ of $7.1~\mathrm{pA}$ and
$4.8~\mathrm{pA}$ respectively. The maximum current difference $\Delta
I_{V_{1}\leftrightarrow T_{-}}$ in Fig.5, top panel, corresponds to
$129~\mathrm{fA}$\ for a background of $8~\mathrm{pA}$. Therefore, the
features described in this article seem measurable experimentally.

\section{VII. Discussion on the spectroscopic entanglement detection scheme}

The present section presents further examination and modifications of the
model used above, in order to put the results of section VI into perspective.

\subsection{VII.1 Use of a more general spin/orbit coupling term}

The $H_{RF}^{so}$ coupling term of Eq.(\ref{hso}) accounts for the coupling
between the CPS and microwave excitations mediated by spin-orbit coupling. The
above sections have used the particular form $\alpha_{i\tau\sigma}%
=\mathbf{i}\sigma\alpha_{i}$ with $\alpha_{i}\in\mathbb{R}$, obtained from a
microscopic description of spin-orbit coupling in a zigzag nanotube quantum
dot\cite{Epaps}. This section discusses the generalization of the results to a
more general coupling $\alpha_{i\tau\sigma}$. Since $H_{RF}^{so}$ must be
hermitian, one can use $\alpha_{i\tau\uparrow}=\alpha_{i\tau}$ and
$\alpha_{i\tau\downarrow}=\alpha_{i\tau}^{\ast}$ without any loss of
generality. The parameter
\begin{equation}
\left\vert \Delta\alpha\right\vert e^{\mathbf{i\varphi}_{\alpha}}=(\alpha
_{LK}+\alpha_{LK^{\prime}})v_{ac}^{L}-(\alpha_{RK}+\alpha_{RK^{\prime}}%
)v_{ac}^{R}\label{sub}%
\end{equation}
with $\mathbf{\varphi}_{\alpha}\in]-\pi,\pi]$ plays a crucial role in this
case. It is convenient to redefine the states $\left\vert T_{+}\right\rangle $
and $\left\vert T_{-}\right\rangle $ more generally as%

\begin{equation}
\left|  T_{+}\right\rangle =\mathbf{i}\mathrm{sgn}(\mathbf{\varphi}_{\alpha
})\left(  e^{-\mathbf{i\varphi}_{\alpha}}\left|  \tilde{T}_{\uparrow
}\right\rangle +e^{\mathbf{i\varphi}_{\alpha}}\left|  \tilde{T}_{\downarrow
}\right\rangle \right)  /\sqrt{2} \label{Tpgene}%
\end{equation}
and
\begin{equation}
\left|  T_{-}\right\rangle =\mathbf{i}\mathrm{sgn}(\mathbf{\varphi}_{\alpha
})\left(  e^{-\mathbf{i\varphi}_{\alpha}}\left|  \tilde{T}_{\uparrow
}\right\rangle -e^{\mathbf{i\varphi}_{\alpha}}\left|  \tilde{T}_{\downarrow
}\right\rangle \right)  /\sqrt{2} \label{Tmgene}%
\end{equation}
with
\begin{align}
\left|  \tilde{T}_{\sigma}\right\rangle  &  =\frac{1}{2}\left(  \sigma
\frac{\Delta_{so}}{\Delta_{r}}-1\right)  \left|  K\sigma,K\sigma\right\rangle
\\
&  -\frac{1}{2}\left(  1+\sigma\frac{\Delta_{so}}{\Delta_{r}}\right)  \left|
K^{\prime}\sigma,K^{\prime}\sigma\right\rangle \nonumber\\
&  +\frac{\Delta_{K/K^{\prime}}}{2\Delta_{r}}\left(  \left|  K\sigma
,K^{\prime}\sigma\right\rangle +\left|  K^{\prime}\sigma,K\sigma\right\rangle
\right) \nonumber
\end{align}
Note that $\left|  T_{+}\right\rangle $ and $\left|  T_{-}\right\rangle $ are
still eigenstates of the hamiltonian $H_{DQD}^{eff}$, with energy
$\delta-2\Delta_{r}$, corresponding to generalized spin triplet states. The
definitions of the other states $\left|  V_{1(2)}\right\rangle $ and $\left|
T_{0}\right\rangle $ remain unchanged. Using expressions (\ref{Tpgene}) and
(\ref{Tmgene}), one obtains
\begin{equation}
\left\langle T_{+}\right|  H_{RF}^{so}\left|  V_{1(2)}\right\rangle =0
\end{equation}
and%
\begin{align}
&  \left\langle T_{_{-}}\right|  H_{RF}^{so}\left|  V_{j}\right\rangle
\label{deltaaplha}\\
&  =-\mathbf{i}ev_{j}\frac{\Delta_{K\leftrightarrow K^{\prime}}}{2\Delta_{r}%
}\left|  \Delta\alpha\right|  \mathrm{sgn}(\mathbf{\varphi}_{\alpha}%
)\cos(\omega_{RF}t)\nonumber
\end{align}
for $j\in\{1,2\}$. In sections II to VI, one uses $\alpha_{i\tau}%
=\mathbf{i}\alpha_{i}$ thus $\varphi_{\alpha}=\mathrm{sgn}(\alpha_{L}%
-\alpha_{R})\pi/2$ and $\left|  T_{\pm}\right\rangle =(\left|  \tilde
{T}_{\uparrow}\right\rangle \mp\left|  \tilde{T}_{\downarrow}\right\rangle
)/\sqrt{2}$ which is in agreement with Eqs.(\ref{Tplus}) and (\ref{Tmoins}).
In this limit, Eq. (\ref{deltaaplha}) agrees with Eq. (\ref{alpha1}).
Equations (\ref{deltaaplha}) and (\ref{sub}) show that even with a more
general coupling term $H_{RF}^{so}$, the matrix elements $\left\langle
T_{_{-}}\right|  H_{RF}^{so}\left|  V_{1(2)}\right\rangle $ still present a
subradiant form. Hence, the entanglement detection scheme discussed in this
article appears to be quite general. Using a more general $H_{RF}^{so}$ will
modify only quantitatively the predictions of section VI.

\subsection{VII.2 Role of $\Delta_{K\leftrightarrow K^{\prime}}\neq0$}

Remarkably, the subradiant matrix elements (\ref{alpha1}) and
(\ref{deltaaplha}) vanish for $\Delta_{K\leftrightarrow K^{\prime}}=0$. The
aim of the present section is to show that using a finite $\Delta
_{K\leftrightarrow K^{\prime}}$ does not represent a fundamental constraint to
have the subradiance effect. Indeed, $\left\vert V_{1(2)}\right\rangle $ can
still be coupled to other triplet states outside of the subspace $\mathcal{E}$
when $\Delta_{K\leftrightarrow K^{\prime}}=0$. This fact is illustrated below,
using $\alpha_{i\tau}=\mathbf{i}\alpha_{i}$ for simplicity. In this case,
$\left\vert V_{1(2)}\right\rangle $ is coupled to a single triplet eigenstate
$\left\vert T_{b}\right\rangle $ of $H_{DQD}^{eff}$ outside the subspace
$\mathcal{E}$, defined by%
\begin{equation}
\left\vert T_{b}\right\rangle =\frac{\alpha_{-}\left(  \left\vert \tilde
{T}_{1\uparrow}\right\rangle -\left\vert \tilde{T}_{2\downarrow}\right\rangle
\right)  }{\sqrt{2(\alpha_{-}^{2}+\alpha_{+}^{2})}}-\frac{\alpha_{+}\left(
\left\vert \tilde{T}_{2\uparrow}\right\rangle -\left\vert \tilde
{T}_{1\downarrow}\right\rangle \right)  }{\sqrt{2(\alpha_{-}^{2}+\alpha
_{+}^{2})}}%
\end{equation}
with%
\begin{equation}
\left\vert \tilde{T}_{1\sigma}\right\rangle =\frac{\Delta_{K/K^{\prime}}%
}{2\tilde{\Delta}_{r}}\left(  \left\vert K^{\prime}\sigma,K^{\prime}%
\sigma\right\rangle -\left\vert K\sigma,K\sigma\right\rangle \right)
+\sigma\frac{\Delta_{so}}{\tilde{\Delta}_{r}}\left\vert K^{\prime}%
\sigma,K\sigma\right\rangle
\end{equation}%
\begin{equation}
\left\vert \tilde{T}_{2\sigma}\right\rangle =\frac{\Delta_{K/K^{\prime}}%
}{2\tilde{\Delta}_{r}}\left(  \left\vert K^{\prime}\sigma,K^{\prime}%
\sigma\right\rangle -\left\vert K\sigma,K\sigma\right\rangle \right)
+\sigma\frac{\Delta_{so}}{\tilde{\Delta}_{r}}\left\vert K\sigma,K^{\prime
}\sigma\right\rangle
\end{equation}%
\begin{equation}
\alpha_{\pm}=\tilde{\alpha}_{L}-\tilde{\alpha}_{R}\pm\frac{\Delta_{so}}%
{\Delta_{r}}(\tilde{\alpha}_{L}+\tilde{\alpha}_{R})
\end{equation}
and
\begin{equation}
\tilde{\alpha}_{L(R)}=\alpha_{L(R)}v_{ac}^{L(R)}%
\end{equation}
such that
\begin{equation}
H_{DQD}^{eff}\left\vert T_{b}\right\rangle =\delta\left\vert T_{b}%
\right\rangle
\end{equation}
One can check:
\begin{align}
\left\langle T_{b}\right\vert h_{so}\left\vert V_{1(2)}\right\rangle  &
=\mathbf{i}ev_{1(2)}\frac{\Delta_{so}^{2}}{\Delta_{r}}(\tilde{\alpha}_{R}%
^{2}-\tilde{\alpha}_{L}^{2})\cos(\omega_{RF}t)\label{c2}\\
&  \times\sqrt{\Delta_{so}^{2}(\tilde{\alpha}_{L}^{2}+\tilde{\alpha}_{R}%
^{2})+\frac{\Delta_{K/K^{\prime}}^{2}}{2}(\tilde{\alpha}_{L}-\tilde{\alpha
}_{R})^{2}}\nonumber
\end{align}
For $\Delta_{K\leftrightarrow K^{\prime}}\rightarrow0$, one finds:%
\begin{equation}
\left\langle T_{b}\right\vert h_{so}\left\vert V_{1(2)}\right\rangle
=\mathbf{i}e\frac{v_{1(2)}(\tilde{\alpha}_{L}^{2}-\tilde{\alpha}_{R}^{2}%
)}{\sqrt{\tilde{\alpha}_{L}^{2}+\tilde{\alpha}_{R}^{2}}}\cos(\omega_{RF}t)
\end{equation}
The coupling between $\left\vert V_{1(2)}\right\rangle $ and $\left\vert
T_{b}\right\rangle $ is subradiant since it vanishes for $\alpha_{R}v_{ac}%
^{R}=\alpha_{L}v_{ac}^{L}$. Nevertheless, for realistic parameters and in
particular $\Delta_{r}\gg t_{eh}$, the transition frequencies $\omega
_{T_{b}V_{1}}$ and $\omega_{T_{b}V_{1}}$ correspond approximately to
$2\Delta_{r}/\hbar$, which is too high for current microwave technology. This
is why this paper focuses on microwave-induced transitions inside the subspace
$\mathcal{E}$.

\subsection{VII.3 Microwave-induced transitions inside the singly occupied
charge sector}

The different eigenstates of $H_{DQD}^{eff}$ in the singly occupied charge
sector can be defined as:
\begin{align}
\left\vert b_{1\sigma}\right\rangle  &  =\frac{1}{2}\sqrt{1-\sigma\frac
{\Delta_{so}}{\Delta_{r}}}\left(  \left\vert K\sigma,0\right\rangle
-\left\vert 0,K\sigma\right\rangle \right) \\
&  +\frac{\Delta_{K/K^{\prime}}}{2\Delta_{r}\sqrt{1-\sigma\frac{\Delta_{so}%
}{\Delta_{r}}}}\left(  \left\vert 0,K^{\prime}\sigma\right\rangle -\left\vert
K^{\prime}\sigma,0\right\rangle \right) \nonumber
\end{align}%
\begin{align}
\left\vert a_{1\sigma}\right\rangle  &  =-\frac{1}{2}\sqrt{1-\sigma
\frac{\Delta_{so}}{\Delta_{r}}}\left(  \left\vert K\sigma,0\right\rangle
+\left\vert 0,K\sigma\right\rangle \right) \\
&  +\frac{\Delta_{K/K^{\prime}}}{2\Delta_{r}\sqrt{1-\sigma\frac{\Delta_{so}%
}{\Delta_{r}}}}\left(  \left\vert 0,K^{\prime}\sigma\right\rangle +\left\vert
K^{\prime}\sigma,0\right\rangle \right) \nonumber
\end{align}%
\begin{align}
\left\vert b_{2\sigma}\right\rangle  &  =-\frac{1}{2}\sqrt{1+\sigma
\frac{\Delta_{so}}{\Delta_{r}}}\left(  \left\vert K\sigma,0\right\rangle
-\left\vert 0,K\sigma\right\rangle \right) \\
&  +\frac{\Delta_{K/K^{\prime}}}{2\Delta_{r}\sqrt{1+\sigma\frac{\Delta_{so}%
}{\Delta_{r}}}}\left(  \left\vert 0,K^{\prime}\sigma\right\rangle
\}-\left\vert K^{\prime}\sigma,0\right\rangle \right) \nonumber
\end{align}
and
\begin{align}
\left\vert a_{2\sigma}\right\rangle  &  =\frac{1}{2}\sqrt{1+\sigma\frac
{\Delta_{so}}{\Delta_{r}}}\left(  \left\vert K\sigma,0\right\rangle
+\left\vert 0,K\sigma\right\rangle \right) \\
&  +\frac{\Delta_{K/K^{\prime}}}{2\Delta_{r}\sqrt{1+\sigma\frac{\Delta_{so}%
}{\Delta_{r}}}}\left(  \left\vert 0,K^{\prime}\sigma\right\rangle
\}+\left\vert K^{\prime}\sigma,0\right\rangle \right) \nonumber
\end{align}
for $\sigma\in\{\uparrow,\downarrow\}$. These states have eigenenergies
$\varepsilon_{1}^{b}$, $\varepsilon_{1}^{a}$, $\varepsilon_{2}^{b}$, and
$\varepsilon_{2}^{a}$ respectively, with
\begin{equation}
\varepsilon_{i}^{b}=\varepsilon-t_{ee}+(-1)^{i}\Delta_{r}%
\end{equation}
and%
\begin{equation}
\varepsilon_{i}^{a}=\varepsilon+t_{ee}+(-1)^{i}\Delta_{r}%
\end{equation}
for $i\in\{1,2\}$. The states $\left\vert b_{1\sigma}\right\rangle $ and
$\left\vert b_{2\sigma}\right\rangle $ can be seen as generalized bonding
states and $\left\vert a_{1\sigma}\right\rangle $ and $\left\vert a_{2\sigma
}\right\rangle $ as generalized antibonding states. This section uses
$\alpha_{i\tau}=\mathbf{i}\alpha_{i}$ for simplicity. The term $H_{RF}^{so}$
couples $\left\vert b_{i\sigma}\right\rangle $ and $\left\vert a_{i\sigma
}\right\rangle $ to $\left\vert b_{i\overline{\sigma}}\right\rangle $ and
$\left\vert a_{i\overline{\sigma}}\right\rangle $ only, for $i\in\{1,2\}$.
Only the transitions $\left\vert a_{i\sigma}\right\rangle \leftrightarrow
\left\vert b_{i\overline{\sigma}}\right\rangle $ correspond to a finite
frequency, i.e. $\omega_{a_{i\sigma}b_{i\overline{\sigma}}}=2t_{ee}/\hbar$.
One can check that%
\begin{equation}
\left\langle b_{i\overline{\sigma}}\right\vert H_{RF}^{so}\left\vert
a_{i\sigma}\right\rangle =-\mathbf{i}e(\alpha_{L}v_{ac}^{L}-\alpha_{R}%
v_{ac}^{R})\cos(\omega_{RF}t)/2 \label{st1}%
\end{equation}
whereas
\begin{align}
\left\langle b_{i\overline{\sigma}}\right\vert H_{RF}^{so}\left\vert
b_{i\sigma}\right\rangle  &  =\left\langle a_{i\overline{\sigma}}\right\vert
H_{RF}^{so}\left\vert a_{i\sigma}\right\rangle \label{st2}\\
&  =\mathbf{i}e(\alpha_{L}v_{ac}^{L}+\alpha_{R}v_{ac}^{R})\cos(\omega
_{RF}t)/2\nonumber
\end{align}
Importantly, the matrix element of Eq.(\ref{st1}) has a subradiant structure.
This property is due to the fact that the states $\left\vert b_{i\overline
{\sigma}}\right\rangle $ and $\left\vert a_{i\sigma}\right\rangle $ are
entangled states with different symmetries, i.e. $\left\vert b_{i\overline
{\sigma}}\right\rangle $ is an antibonding state which contains some
$\left\vert \tau\overline{\sigma},0\right\rangle +\left\vert 0,\tau
\overline{\sigma}\right\rangle $ components whereas $\left\vert a_{i\sigma
}\right\rangle $ is a bonding state which contains $\left\vert \tau
\sigma,0\right\rangle -\left\vert 0,\tau\sigma\right\rangle $ components. This
is analogous to the fact that the elements $\left\langle T_{_{-}}\right\vert
H_{RF}^{so}\left\vert V_{j}\right\rangle $ couple a state $\left\vert
V_{j}\right\rangle $ with a spin-singlet component to a spin-triplet state
$\left\vert T_{_{-}}\right\rangle $. In contrast, the matrix elements of
Eq.(\ref{st2}) are not subradiant because they couple two entangled states
with similar symmetries, i.e. two bonding or two antibonding states.

Due to the subradiant form of Eq. (\ref{st1}), the transitions $\left\vert
a_{i\sigma}\right\rangle \leftrightarrow\left\vert b_{i\overline{\sigma}%
}\right\rangle $ can lead to a non-monotonic variation of the CPS input
current as a function of e.g. $v_{ac}^{L}$, due to another type of
entanglement than the one discussed in section VI. Therefore, in the context
of the characterization of split Cooper pairs entanglement, one needs to find
a way to discriminate possible current resonances corresponding to the
transitions $\left\vert a_{i\sigma}\right\rangle \leftrightarrow\left\vert
b_{i\overline{\sigma}}\right\rangle $ and $\left\vert V_{1(2)}\right\rangle
\leftrightarrow\left\langle T_{_{-}}\right\vert $. In practice, this should be
feasible by studying how the different resonance frequencies vary with the DC
gate voltages of the two dots. Indeed, $\omega_{a_{i\sigma}b_{i\overline
{\sigma}}}$ does not depend on the parameter $\delta$, contrarily to
$\omega_{V_{1}T_{-}}$ and $\omega_{T_{-}V_{2}}$. Therefore, possible current
resonances due to $\left\vert a_{i\sigma}\right\rangle \leftrightarrow
\left\vert b_{i\overline{\sigma}}\right\rangle $ transitions should appear as
horizontal lines in Fig. 4.b. This effect was nevertheless disregarded in
section VI, assuming that $\omega_{a_{i\sigma}b_{i\overline{\sigma}}}$ is too
large to be accessible experimentally. Studying quantitatively the possibility
to observe the resonances $\left\vert a_{i\sigma}\right\rangle \leftrightarrow
\left\vert b_{i\overline{\sigma}}\right\rangle $ requires to go beyond the
approximation of an electronic tunnel rate $\Gamma_{N}$ to the normal leads
which is independent from the dot orbital and spin indices \cite{noteb}.

\subsection{VII.4 Simplified model without the K/K' degeneracy}

It is interesting to discuss a model without the K/K' degree of freedom to
show simply how the subradiance property arises.

\subsubsection{Case of coherent Cooper pair injection}

Let us assume that each of the two CPS dots has a single orbital. One can note
$\left\vert \sigma,\sigma^{\prime}\right\rangle $ a CPS doubly occupied state
with a spin $\sigma(\sigma^{\prime})$ on dot $L(R)$. In the case of coherent
Cooper pair injection, the double quantum dot effective hamiltonian can be
written\cite{Eldridge}%
\begin{align}
H_{DQD}^{eff}  &  =\varepsilon(n_{L\sigma}+n_{R\sigma})\label{Hsimple}\\
&  +(t_{eh}/\sqrt{2})\left(  d_{L\uparrow}^{\dag}d_{R\downarrow}^{\dag
}-d_{L\downarrow}^{\dag}d_{R\uparrow}^{\dag}+h.c.\right)  +H_{int}\nonumber
\end{align}
where $H_{int}$ still forbids the double occupation of each dot. One uses
above $n_{i\sigma}=d_{i\sigma}^{\dag}d_{i\sigma}$ with $d_{i\sigma}^{\dag}$
the creation operator for an electron with spin $\sigma$ in dot $i\in\{L,R\}$.
Let us furthermore assume that there also exists a spin-flip coupling term to
the microwave signal, with the form%
\begin{align}
H_{RF}^{sf}  &  =-%
{\displaystyle\sum\limits_{i}}
\alpha_{i}ev_{ac}\cos(\omega_{RF}t)(d_{i\uparrow}^{\dag}d_{i\downarrow
}+d_{i\downarrow}^{\dag}d_{i\uparrow})\nonumber\\
&  =%
{\displaystyle\sum\limits_{i}}
\lambda_{i}(d_{i\uparrow}^{\dag}d_{i\downarrow}+d_{i\downarrow}^{\dag
}d_{i\uparrow})
\end{align}
Such a spin-flip coupling can be due for instance to the magnetic field
associated with the microwave irradiation. In practice, this term should have
a very weak amplitude, but it is nevertheless discussed here for fundamental purposes.

The term in $t_{eh}$ hybridizes the CPS empty state $\left\vert
0,0\right\rangle $ with the singlet state $\left\vert \mathcal{\tilde{S}%
}\right\rangle =(\left\vert \uparrow,\downarrow\right\rangle -\left\vert
\downarrow,\uparrow\right\rangle )/\sqrt{2}$, so that an anticrossing appears
again in the spectrum of the CPS even-charged states. For simplicity, it is
asssumed below that the double occupation energy $\delta=2\varepsilon$ of the
CPS is degenerate with the energy of $\left\vert 0,0\right\rangle $, i.e.
$\delta=0$. In this case, one can use the orthonormalized basis $\mathcal{A}%
=\{\tilde{V}_{1},\tilde{V}_{2},\left\vert \tilde{T}_{a}\right\rangle
,\left\vert \tilde{T}_{b}\right\rangle ,\left\vert \tilde{T}_{0}\right\rangle
\}$ of eigenstates of (\ref{Hsimple}) in the even charge sector , with%

\begin{equation}
\left\vert \tilde{V}_{1(2)}\right\rangle =\left(  \left\vert 0,0\right\rangle
\pm\left\vert \mathcal{\tilde{S}}\right\rangle \right)  /\sqrt{2} \label{11}%
\end{equation}%
\begin{equation}
\left\vert \tilde{T}_{a(b)}\right\rangle =\left(  \left\vert \uparrow
,\uparrow\right\rangle \pm\left\vert \downarrow,\downarrow\right\rangle
\right)  /\sqrt{2} \label{22}%
\end{equation}
and%
\begin{equation}
\left\vert \tilde{T}_{0}\right\rangle =\left(  \left\vert \uparrow
,\downarrow\right\rangle +\left\vert \downarrow,\uparrow\right\rangle \right)
/\sqrt{2} \label{33}%
\end{equation}
The states $\left\vert \tilde{V}_{1}\right\rangle $ and $\left\vert \tilde
{V}_{2}\right\rangle $ have energies $\tilde{E}_{1}$ and $\tilde{E}_{2}$ given
by $\tilde{E}_{1(2)}=\pm t_{eh}$. They play the role of the states $\left\vert
V_{1}\right\rangle $ and $\left\vert V_{2}\right\rangle $ of section VI. It is
convenient to define $\left\vert \tilde{T}_{a}\right\rangle $ and $\left\vert
\tilde{T}_{b}\right\rangle $ as superpositions of triplet states with equal
spins. The states $\left\vert \tilde{T}_{0}\right\rangle $, $\left\vert
\tilde{T}_{a}\right\rangle $ and $\left\vert \tilde{T}_{b}\right\rangle $ have
an energy $\delta=0$. One can check straigthforwardly that
\[
H_{RF}^{sf}\left\vert \mathcal{\tilde{S}}\right\rangle =(\lambda_{R}%
-\lambda_{L})\left\vert \tilde{T}_{b}\right\rangle
\]
thus%
\begin{equation}
\left\langle \tilde{T}_{b}\right\vert H_{RF}^{sf}\left\vert \tilde{V}%
_{1(2)}\right\rangle =\pm(\lambda_{R}-\lambda_{L})/2 \label{subb}%
\end{equation}
whereas $\left\langle \tilde{T}_{a(0)}\right\vert H_{RF}^{sf}\left\vert
\tilde{V}_{1(2)}\right\rangle =0$ and $\left\langle \tilde{V}_{2}\right\vert
H_{RF}^{sf}\left\vert \tilde{V}_{1}\right\rangle =0$. The states $\tilde
{V}_{1(2)}$ are thus coupled by $H_{RF}^{sf}$ to a single state $\left\vert
\tilde{T}_{b}\right\rangle $, with a subradiant matrix element (\ref{subb}).
This illustrates the universality of the mechanism discussed in section VI.

\subsubsection{Case of incoherent singlet injection}

One can model na\"{\i}vely the incoherent injection of Cooper pairs inside the
CPS by assuming that up spins are always injected inside the left dot and
right spins inside the right dot. This requires to replace the hamiltonian
(\ref{Hsimple}) by%
\begin{align}
H_{DQD}^{eff} &  =\varepsilon(n_{L\sigma}+n_{R\sigma})\label{Hincoh}\\
&  +t_{eh}(d_{L\uparrow}^{\dag}d_{R\downarrow}^{\dag}+d_{R\downarrow
}d_{L\uparrow})+H_{int}\nonumber
\end{align}
One can use again $\delta=2\varepsilon=0$ for simplicity.\ In this case, one
can define an orthonormalized basis $\mathcal{B}=\{W_{1},W_{2},\left\vert
\tilde{T}_{c}\right\rangle ,\left\vert \tilde{T}_{d}\right\rangle ,\left\vert
\tilde{T}_{e}\right\rangle \}$ of eigenstates of (\ref{Hincoh}) in the even
charge sector, with%
\begin{equation}
\left\vert W_{1(2)}\right\rangle =\left(  \left\vert 0,0\right\rangle
\pm\left\vert \uparrow,\downarrow\right\rangle \right)  /\sqrt{2}%
\end{equation}%
\begin{equation}
\left\vert \tilde{T}_{c}\right\rangle =\left(  \lambda_{R}\left\vert
\uparrow,\uparrow\right\rangle +\lambda_{L}\left\vert \downarrow
,\downarrow\right\rangle \right)  /\sqrt{\lambda_{L}^{2}+\lambda_{R}^{2}}%
\end{equation}%
\begin{equation}
\left\vert \tilde{T}_{d}\right\rangle =\left(  \lambda_{L}\left\vert
\uparrow,\uparrow\right\rangle -\lambda_{R}\left\vert \downarrow
,\downarrow\right\rangle \right)  /\sqrt{\lambda_{L}^{2}+\lambda_{R}^{2}}%
\end{equation}
and $\left\vert \tilde{T}_{e}\right\rangle =\left\vert \downarrow
,\uparrow\right\rangle $. The role of the states $\left\vert V_{1}%
\right\rangle $ and $\left\vert V_{2}\right\rangle $ of section VI is now
played by $\left\vert W_{1}\right\rangle $ and $\left\vert W_{2}\right\rangle
$. The states $\left\vert W_{1}\right\rangle $ and $\left\vert W_{2}%
\right\rangle $ have again energies $\tilde{E}_{1}$ and $\tilde{E}_{2}$
defined in the previous section, whereas the states $\left\vert \tilde{T}%
_{c}\right\rangle $, $\left\vert \tilde{T}_{d}\right\rangle $ and $\left\vert
\tilde{T}_{e}\right\rangle $ have an energy $\delta=0$. The states $\left\vert
\tilde{T}_{a}\right\rangle $ and $\left\vert \tilde{T}_{b}\right\rangle $ of
the previous section are still CPS eigenstates, but it is more convenient to
use the eigenstates $\left\vert \tilde{T}_{c}\right\rangle $ and $\left\vert
\tilde{T}_{d}\right\rangle $ to study the effect of $H_{RF}^{sf}$. Due to the
term in $t_{eh}$, the states $\left\vert W_{1}\right\rangle $ and $\left\vert
W_{2}\right\rangle $ still form an anticrossing in the energy spectrum of the
CPS. Hence, such an anticrossing is not characteristic from the injection of
entangled Cooper pairs. The only state of $\mathcal{B}$ connected to
$\left\vert W_{1(2)}\right\rangle $ by $H_{RF}^{sf}$ is $\left\vert \tilde
{T}_{c}\right\rangle $, with a matrix element%
\begin{equation}
\left\langle \tilde{T}_{c}\right\vert H_{RF}^{sf}\left\vert W_{1(2)}%
\right\rangle =\pm\sqrt{2(\lambda_{L}^{2}+\lambda_{R}^{2})}%
\end{equation}
which is not subradiant, but increases monotonically with $\lambda_{R}$ and
$\lambda_{L}$. Therefore, the subradiance property is lost when Cooper pairs
are injected inside the CPS in a product state instead of an entangled state.
Similar results are expected for a model including the $K/K^{\prime}$ degree
of freedom.This illustrates that the subradiance property is a good indication
of the injection of entangled Cooper pairs inside the CPS. More sophisticated
descriptions of incoherent injection of Cooper pairs into the CPS are beyond
the scope of this article.

\section{VIII. Comparison with an alternative setup: the CPS embedded in a
microwave cavity}

This section compares the experimental scheme proposed in Ref.\cite{CKLY} to
the scheme discussed in the present article. Reference \cite{CKLY} suggests to
observe the minus sign in Equation (\ref{alpha1}) by inserting the CPS inside
a coplanar microwave cavity to obtain a lasing effect involving the
$\left\vert V_{1}\right\rangle \rightarrow\left\vert T_{-}\right\rangle $
transition. The minus sign in Equation (\ref{alpha1}) leads to a non-monotonic
dependence of the number of photons in the cavity as a function of the
coefficients $\alpha_{L}$ and $\alpha_{R}$, which mediate a coupling between
the CPS and the electric field conveyed by the cavity. Since this electric
field can be considered as constant over the whole CPS area, it is necessary
to be able to vary $\alpha_{L}$ independently from $\alpha_{R}$ to observe a
non-monotonic behavior in the number of photons. This can require to
complexify the CPS design, for instance. In the present scheme, such a control
on $\alpha_{L}$ and $\alpha_{R}$ is not necessary since it is sufficient to
vary independently the amplitudes $v_{ac}^{L}$ and $v_{ac}^{R}$. This can be
naturally achieved by using two independent microwave supplies for the two
gates. The advantage of the scheme presented in Ref.\cite{CKLY} is that the
signal to be measured is a large photon number which can be obtained by
measuring the power spectrum emitted by the cavity. In other words, the scheme
of Ref.\cite{CKLY} exploits the fact that the lasing effect provides an
intrinsic amplification process for the $V_{1}\leftrightarrow T_{-}$
transitions. In the present scheme, the measurement seems a bit more difficult
since the current peaks to be measured are very small. Nevertheless, such
current amplitudes are measurable, in principle\cite{Meyer}. Therefore, the
scheme presented in this reference could be an interesting alternative
approach to demonstrate the coherent injection of singlet Cooper pairs inside
a CPS. This scheme furthermore allows to study also the $\left\vert
V_{2}\right\rangle \leftrightarrow\left\langle T_{_{-}}\right\vert $
transition, which is not possible with the scheme of Ref. \cite{CKLY}.

\section{IX. Conclusion}

The DC current response of a double-quantum-dot based Cooper pair beam
splitter (CPS) to a microwave gate irradiation is a very rich source of
information on Cooper pair splitting. In particular, it can reveal the
entanglement of spin-singlet Cooper pairs injected inside the CPS. This
article illustrates this property for a double quantum dot formed inside a
carbon nanotube with typical parameters. If they are spin-entangled, the
injected pairs are coupled to other CPS states through some subradiant
microwave transitions mediated by spin-orbit coupling. This property can be
revealed by applying to the two CPS quantum dots two on-phase microwave gate
voltages. The spin-orbit mediated microwave transitions cause DC current
resonances at the input of the CPS. The subradiance property manifests in a
strongly nonmonotonic variation of these current resonances with the amplitude
of the microwave signal applied to one of the two CPS dots. This behavior does
not depend on details of the model like the exact form of the spin-orbit
interaction term. Similarly, the presence of atomic disorder in the nanotube
has to be assumed only for quantitative reasons. More generally, the
entanglement detection scheme discussed in this work could be generalized to
other types of quantum dots with spin-orbit coupling like e.g. InAs quantum
dots, in principle. For simplicity, this article discusses the limit where the
intra-dot charging energies are very strong, so that there cannot be two
electrons at the same time on the same dot. For smaller charging energies, the
efficiency of Cooper pair splitting should be decreased. Nevertheless, if the
CPS produces entangled split Cooper pairs with a sufficient rate, the
resulting subradiant current peaks should still be observable. Interestingly,
the bonding or antibonding single particle states delocalized on the two dots
of the CPS can also cause a subradiant current resonance, because they present
another type of entanglement. However, in principle, this resonance can easily
be discriminated from the subradiant resonances caused by split Cooper pairs,
because of a different dependence on the CPS DC gate voltages.

\textit{I acknowledge useful discussions with T. Kontos, J. Viennot and A.
Levy Yeyati. This work has been financed by the EU-FP7 project SE2ND[271554].}


\begin{thebibliography}{99}
\bibitem {noteb}Note that in principle, the term $H_{RF}^{g}$ can induce
transitions $\left\vert a_{i\sigma}\right\rangle \leftrightarrow\left\vert
b_{i\sigma}\right\rangle $, at the frequency $2t_{ee}/\hbar$. However, the
states $\left\vert a_{i\sigma}\right\rangle $ and $\left\vert b_{i\sigma
}\right\rangle $ have similar weights in $\left\vert K\sigma,0\right\rangle $,
$\left\vert 0,K\sigma\right\rangle $, $\left\vert K^{\prime}\sigma
,0\right\rangle $, and $\left\vert 0,K^{\prime}\sigma\right\rangle $.
Therefore, these transitions should not be visible in the CPS input current.
\end{thebibliography}
\end{document}